\documentclass[a4paper,12pt]{article}
\usepackage{
latexsym,amssymb}
\usepackage{graphicx}
\usepackage{cite}
\global\arraycolsep=2pt 
\input{epsf}

\def\eq#1\en{\begin{equation}#1\end{equation}}
\def\s[#1,#2]{[#1\stackrel{{\displaystyle\star}}{,}#2]}
\def\pp#1{\partial_#1}

\newcommand{\eqa}{\begin{eqnarray}}
\newcommand{\ena}{\end{eqnarray}}
\newcommand{\enn}{\nonumber \end{equation}}

\def\sk{\vskip .4cm}
\def\noi{\noindent}

\def\QEDp{{{\rm{QED}}_+}}
\def\QEDm{{{\rm{QED}}_-}}
\def\SW{{{\mbox {\small {SW}}}}}
\def\P*{{\Psi^*}}
\def\PL*{{\Psi_{\!L}^{*}}}
\def\PR*{{\Psi_{\!R}^{*}}}
\def\p*{{\psi^*}}
\def\U{\Upsilon}
\def\de{\delta}
\def\L{{\Lambda}}
\def\lh{{\widehat \Lambda}}

\def\*{\star}
\def\m*{{\star^{\mbox{${_{^{{\!{o{^{_{^{_{\!\!}}}}}p}}}}}$}}}}

\def\psih{\widehat \psi}
\def\Psih{\widehat \Psi}

\def\Fh{\widehat F}
\def\Ah{\widehat A}

\def\D/h{\widehat{\fmslash D}}
\def\Sh{\widehat S}
\def\ell{{{\it l}}}
\def\Tcal{{\cal{T}}}
\def\1{1\!\!^{\!}1}

\def\Tr{{\rm{Tr}}}
\def\dim{{\rm{d}}}
\def\Lambdah{\widehat{\Lambda}}
\def\th{\theta}

\def\al{\alpha}

\def\be{\beta}

\def\bx{\mbox{\boldmath ${x}$}}
\def\bp{\mbox{\boldmath ${p}$}}

\def\5bar{{\overline 5}}
\def\pu{{\!\cdot^{\!}}}
\def\sma#1{\mbox{\footnotesize #1}}

\def\sr{\sigma_{_{\!}\Psi_{\!R}}}

\begin{document}
\makeatletter
\def\fmslash{\@ifnextchar[{\fmsl@sh}{\fmsl@sh[0mu]}}
\def\fmsl@sh[#1]#2{%
  \mathchoice
    {\@fmsl@sh\displaystyle{#1}{#2}}%
    {\@fmsl@sh\textstyle{#1}{#2}}%
    {\@fmsl@sh\scriptstyle{#1}{#2}}%
    {\@fmsl@sh\scriptscriptstyle{#1}{#2}}}
\def\@fmsl@sh#1#2#3{\m@th\ooalign{$\hfil#1\mkern#2/\hfil$\crcr$#1#3$}}
\makeatother
%
\thispagestyle{empty}
\begin{titlepage}

\begin{flushright}
LMU-TPW 2002-02\\
May 2002
\end{flushright}

\vspace{0.3cm}
\boldmath
\begin{center}
  {\Large {\bf $\!\!$Non-Commutative GUTs, Standard Model and 
$C,_{\!}P,T\!\!\!\!$}}
\end{center}
\vspace{-0.4cm}
\unboldmath
\vspace{0.8cm}
\begin{center}
{
{{\bf P. Aschieri${}^{1}$, B.\ Jur\v co${}^{1}$, 
P.\ Schupp${}^{2}$, J.\ Wess${}^{1,3}$}}}
\end{center}
\vskip 1em
\begin{center}
${}^{1}$Sektion Physik, Universit\"at M\"unchen,\\
  Theresienstra{\ss}e 37, 80333 M\"unchen, Germany\\
${}^{2}$International University Bremen,\\ School of Engineering and Science,\\
Campus Ring 1, 28759 Bremen, Germany\\
${}^{3}$Max-Planck-Institut f\"ur Physik\\
        F\"ohringer Ring 6, 80805 M\"unchen, Germany\\[1em]
 \end{center}

\vspace{\fill}

\begin{abstract}
\noindent 
Noncommutative Yang-Mills theories
are sensitive to the choice of the 
representation that enters 
in the gauge kinetic term.
We constrain this ambiguity by considering grand unified
theories.
We find that at first order in the noncommutativity 
parameter $\th$,  $SU(5)$ is not truly a unified theory, while 
$SO(10)$ has a unique noncommutative
generalization. In view of these results we discuss the noncommutative
SM theory that is compatible with $SO(10)$ GUT and
find that there are no modifications to the  SM gauge kinetic term 
at lowest order in $\th$\,.

We study in detail the reality, hermiticity and $C,P,T$ 
properties of the Seiberg-Witten map 
and of the resulting effective actions expanded in ordinary fields. 
We find that in models of GUTs (or compatible with GUTs) right-handed 
fermions and left-handed ones appear with opposite Seiberg-Witten map.
\end{abstract}
\end{titlepage}

\section{Introduction}
Noncommutative (NC) space-time and in general 
noncommutative geometry seem a natural arena where
to study a quantum theory of general relativity; however 
one does not need to invoke quantum gravity to 
motivate noncommutative space-time models. In M-theory and 
in open string theory, in the presence of a nonvanishing NS $B$ field,
the effective physics on D-branes can be described by a noncommutative 
gauge theory \cite{Connes:1997cr}. Here the source of noncommutativity
is the two-form $B$.
The easiest and most studied example is the case of
constant $B$, this induces the following noncommutativity
$\s[x^\mu,x^\nu]\equiv x^\mu\*x^\nu-x^\nu\* x^\mu=i\th^{\mu\nu}$ 
with $\th^{\mu\nu}$ constant (and depending on $B$).
$\theta^{\mu\nu}$ fixes directions in space-time. With respect to fixed
$\theta^{\mu\nu}$ 
we see that the Lorentz group is broken in a spontaneous way; 
in a bigger theory where the $B$ field is dynamical and not frozen 
to a constant value we have Lorentz  covariance. Also, at low
energies ($E^2\th^{\mu\nu}\ll 1$) Lorentz symmetry should be recovered.
The product $\*$ is the Moyal star product. On functions $f,g$ 
we have $f\*g=f {_{\,}}e^{{i\over 2}{\th}^{\mu\nu}{\stackrel{\leftarrow}
{\partial}_{\!\mu}}{\stackrel{\rightarrow}{\partial}_{\!\nu}}} {_{\,}}g_{_{_{_{_{}}}}}$.

Recently there has been a lot of interest in the study of realistic 
${}^{^{{^{}}}}$particle models based on the $\s[x^\mu,x^\nu]=i\th^{\mu\nu}$ 
noncommutative space-time \cite{Jabbari,Chu,SM}. The general idea
being that a noncommutative space-time structure is not necessarily 
a Planck length phenomenon.
Bounds on the noncommutative scale from
collider physics can be as low as a
few TeV's~\cite{ccphy} and it is therefore interesting to compare the 
theoretical predictions of these models~\cite{ccphy,Behr}
with near future experiments. Bounds from low energy physics,
in particular from clock comparison experiments are
in general much higher~\cite{Carlson:2001sw}. These bounds must be
interpreted with care, however, for several reasons: 1) Due to the
nature of the experiments, the bound concerns the spatial and temporal
average of the noncommutativity tensor $\theta^{\mu\nu}$.
Non-constant (slowly varying)  components of
$\theta$ may not be directly affected by the bound. 2) The bounds are
based on loop calculations in noncommutative field theory
and in particular  in NC QCD. These
theories are so far not well understood as full fledged quantum
theories. There may, e.g., be additional terms in the quantum
action (that are consistent with
the symmetries) and whose coefficients (rather than the overall
noncommutativity scale) are bounded by the experiments.%
\footnote{See also \cite{Carlson:2001bk} for a way to avoid 
the bounds.}
Among the different NC (star product) generalization of the Standard Model (SM) (\cite{Jabbari,SM}),
the one in \cite{SM} is the only one  based directly on the SM gauge group
$U(1)_Y\otimes SU(2)_L\otimes SU(3)_{\textstyle c}$. It also has 
the same particle
content as the ordinary SM. The major new aspect
of \cite{SM}, with respect to the ordinary SM, is the appearence of new 
$\th$-dependent interactions that are dictated by requiring that both 
noncommutative and commutative gauge transformations are a symmetry of 
the action. 
The construction of this model is based on Seiberg-Witten map
(SW map) between commutative and noncommutative gauge 
transformations and fields. SW map was initially introduced 
for $U(N)$ gauge fields in \cite{SW}, in the context of open string 
theory (and the zero slope limit $\alpha'\rightarrow 0$ \cite{SW}). 
It has then been studied in the case of arbitrary gauge groups 
\cite{Madore, Jurco:2001rq, Bonora, Brace, Bars, Dorn } 
and space-time dependent parameters $\th^{\mu\nu}$ \cite{Jurco}.
The SW map and the $\*$ product allow us to expand
the noncommutative action order by order in $\th$ and to express 
it in terms of ordinary commutative fields, 
so that one can then study the property of 
this $\th$-expanded commutative action.
It turns out that given a commutative YM theory, 
SW map and commutative/noncommutative gauge invariance 
are in general not enough in order to single out a unique 
noncommutative generalization of the
original YM theory. One can follow different criteria in order to
select a specific noncommutative generalization. We here focus on a
classical analysis, in particular imposing the constraint that 
the noncommutative generalization of the Standard Model should 
be compatible with noncommutative GUT theories. 
Another issue would be to single out a noncommutative SM or GUT that 
is well behaved at the quantum level. 
We refer to the problems relative to 
renormalization and chiral gauge anomalies. 
For example $\th$-expanded massless QED is not 
renormalizable \cite{Wulkenhaar1}, 
however the photon self energy is renormalizable to
all orders \cite{Grosse}, and, adding just one
extra fermion -- gauge boson interaction,
one obtains one-loop renormalizability of the full theory \cite{Wulkenhaar2}. 
Chiral anomalies in the context of 
$\th$-expanded actions (as far as we know) have not yet been studied
(see however \cite{Ghosh}).
The analysis in the case of $U(N)$ gauge theories 
(no SW map, no $\th$-expansion) shows that chiral theories are not
anomaly free \cite{Martin, Tomasiello}.

\sk
In this  paper we first present a general study of the ambiguities that are
present when constructing NCYM theories. We then see that at first order in 
$\th$ there is no ambiguity in $SO(10)$ NCYM theory. In particular 
no triple gauge bosons coupling  of the kind  $\th FFF$ (indices arbitrarily 
contracted) is present.
We further study the noncommutative SM compatible with $SO(10)$: it is 
constructed using just left handed fermions and antifermions, so that
adding a left handed antineutrino
($\nu^{\scriptscriptstyle  C}_{~L}=
-i\sigma_2{_{_{\,}}}{\nu_{_{\!}R}}^{^{_{_{_{\scriptstyle *}}}}}$)
one obtains the $SO(10)$ chiral fermion multiplet. 
If (as it is natural) one considers just the
adjoint representation $\rho_{\rm{adj.}}^{}$ 
and the fermion representation $\rho_{\rm{f}}^{}$ 
in this SM noncommutative gauge kinetic term 
$\sum_\rho\Tr(\rho(\hat{F})\rho(\hat{F}))$,
then here too no $\th FFF$ term is present.
This is not the case for the NCSM in \cite{SM} because there 
the non-chiral vector 
$\Psi'=(u^i_L,d^i_L\,,\,u^i_R\,d^i_R\,,\,\nu_L,e^-_L\,,\,e^-_R)$
is considered.

We next study the reality, hermiticity, 
charge conjugation, parity and time reversal properties of 
the SW map and of $\th$-expanded NCYM theories. 
This constraints the possible freedom in the choice of a
``good'' SW map.
There are in principle two choices for the result
of the combined charge parity and time reversal 
transformation on the NC algebra (the
star product): the CPT operator maps the star product
to the opposite star product (with $-\theta$ in place
of $\theta$); alternatively one can define a $cpt$ transformation 
that leaves the $\*$ product invariant.
In \cite{Sheikh-Jabbari:2000vi} the $C,P,T$ properties of NCQED
were studied assuming the usual $C,P$ and $T$ transformations
also for noncommutative fields. Here we {\sl show} that the usual
$C,P,T$ transformation on commutative spinors and nonabelian
gauge potentials imply, via SW map, the same $C,P,T$
transformations for the noncommutative spinors and gauge potentials.
We also see that $CPT$ is always a  symmetry of noncommutative actions. 
The $CPT$ operator is compatible with the SW map.
The $cpt$ operator on the other hand maps the SW map
to the opposite SW map. It is also not difficult to construct NCYM actions that are
even under $\th\rightarrow -\th$ and thus invariant under this $cpt$ 
transformation.

The reality property of the SW map is used to 
analyze the difference between the SM in \cite{SM} and the GUT inspired 
SM proposed here. 
It is a basic one, and can be studied also in a QED model. 
While in \cite{SM}, and in general in the literature,
left and right handed components of a noncommutative spinor field are 
built with the same SW map, we here use and advocate a different
choice: if noncommutative left handed fermions are built with the 
$+\th$ SW map then their right handed companions should be built with 
the $-\th$ SW map; this implies that both noncommutative $\psi_L$ and
$\psi^{\scriptscriptstyle C}_{~L}\equiv
-i\sigma_2{_{_{\,}}}{\psi_{_{\!}R}}^{^{_{_{_{\scriptstyle *}}}}}$
are built with the $+\th$ SW map.
In other words, with this choice, noncommutativity does not
distinguish 
between a left handed fermion and a left handed antifermion, but 
does distinguish between fermions with different chirality.
This appears to be the only choice compatible with GUT theories.

\sk
The paper is organized as follows. In Section 2 we construct and 
discuss the ambiguities in noncommutative $SO(10)$, $SU(5)$ and
$U(1)_Y\otimes SU(2)_L\otimes SU(3)_{\textstyle c}$ YM theories 
with fermion matter. We then study the Higgs sector of 
the noncommutative SM and the
Higgs sector of $SO(10)$. In sections 3 and 4 we study the reality,
hermiticity and $C,P,T$ properties of NCYM actions. 
In Subsection 4.1 the difference between
actions built respectively with the $+\th$ and the $-\th$ choice 
for right handed fermions is described. In Section 5
we see that if $\th$ properly transforms under $C,P,T,$ then NCYM
actions have the same $C^{\!}P$ and $T$ symmetries as their corresponding 
commutative ones. The $cpt$ transformation is then considered.
In the Appendix a general expression of the SW map at first order in $\th$
is given; tensor products of noncommutative gauge transformations are
also considered.
\sk

\section{Building NCYM theories}
Consider an ordinary ``commutative'' YM action with  gauge group $G$, 
where $G$ is a compact simple Lie group, and one fermion multiplet $\Psi$
\eq
S=\int d^4x \,{-1\,\over 2 g^2}^{\,} 
\Tr(F_{\mu\nu}F^{\mu\nu})  + 
\overline{\Psi}  i
{\fmslash D} \Psi \label{A1m}
\en
This action is gauge invariant under 
\eq
\delta\Psi=i\rho_\Psi(\Lambda)\Psi\label{psi}
\en
where $\rho_{\Psi}$ is the representation of $G$  determined by the
multiplet $\Psi$.
Following \cite{Jurco:2001rq} the noncommutative generalization of 
(\ref{A1m}) is given by 
\eq
\Sh=\int d^4x \,{-1\,\over 2 g^2}^{\,} 
T^{\!}r(\Fh_{\mu\nu}\*\Fh^{\mu\nu})  + 
\overline{\widehat \Psi} \star i
\widehat{\fmslash D} \widehat \Psi \label{Action1multiplet}
\en
where the noncommutative field strength $\Fh$ is defined by
\eq
\widehat F_{\mu\nu}  = \pp\mu\widehat A_\nu 
- \pp\nu\widehat A_\mu  -i\s[\widehat A_\mu,\widehat A_\nu],
\en
and where both the noncommutative fields $\Ah$ and $\Fh$ are 
hermitian: 
$\Ah_\mu^{\,\dagger}=\Ah_\mu\,,~{\Fh}^{\,\dagger}_{\mu\nu}={\Fh}_{\mu\nu}$. 
The covariant derivative is given by
\eq
\widehat D_\mu \widehat\Psi = \pp\mu \widehat\Psi 
- i \rho_\Psi(\widehat A_\mu)\star\widehat\Psi\,.
\label{covder}
\en
The action (\ref{Action1multiplet}) is invariant under the 
noncommutative gauge transformations 
\eqa
\hat\delta\widehat\Psi = i \rho_\Psi(\widehat\Lambda) \star
\widehat\Psi~~~~~~~~~~~~~~~~~~~~~~~~~~~~~~~~~~~~~~~~~~~~ \label{deltapsi}\\
~~\hat\delta \widehat A_\mu = \pp\mu\widehat\Lambda 
+ i\s[\widehat\Lambda,\widehat A_\mu] ~~~~~
{}~~~\!\!\Rightarrow~~~\hat\delta\widehat F_{\mu\nu} 
= i\s[\widehat\Lambda,\widehat F_{\mu\nu}]~.
\label{deltaA}
\ena

The fields $\Ah$, $\Psih$ and $\lh$ are functions of the commutative 
fields $A,\Psi, \Lambda$ and the noncommutativity parameter $\th$
via the SW map \cite{SW}. 
 At first order in $\th$ we have
(see Section 3 for the freedom in the choice of SW map; see the
appendix for the most general SW map at first order in $\th$)
\begin{eqnarray}
\widehat A_\xi[A,\th] & = & A_\xi 
+ \frac{1}{4} \theta^{\mu\nu}\{A_\nu,\pp\mu A_\xi\} + 
\frac{1}{4} \theta^{\mu\nu}\{F_{\mu\xi},A_\nu\} +  \mathcal{O}(\theta^2)
\label{SWA}\\
\widehat\Lambda[\Lambda, A, \th]  &=&  \Lambda 
+ \frac{1}{4} \theta^{\mu\nu}\{\pp\mu \Lambda , A_\nu\}+ \mathcal{O}(\theta^2)\label{SWLambda}~\\
\widehat \Psi[\Psi,A,\th] & = & \Psi 
+ \frac{1}{2} \theta^{\mu\nu}\rho_\Psi(A_\nu)\pp\mu\Psi
+\frac{i}{8}\theta^{\mu\nu}[\rho_\Psi(A_\mu), \rho_\Psi(A_\nu)] \Psi + \mathcal{O}(\theta^2)
 \label{SWPsi}
\end{eqnarray}
In terms of the commutative fields 
the action (\ref{Action1multiplet}) is also invariant
under the ordinary gauge transformation 
$\delta A_\mu = \pp\mu\Lambda  + i[\Lambda, A_\mu]$,
$\delta\Psi=i\rho_\Psi(\Lambda)\Psi$.
\sk
In (\ref{Action1multiplet}) the information on the gauge group $G$ 
is through the dependence of the noncommutative fields on the 
commutative ones. The commutative
gauge potential $A$ and gauge parameter $\Lambda$ are valued in the
$G$  Lie algebra, $A=A^aT^a, \L =\L^aT^a$. It follows that $\Ah$ and $\Lambdah$ 
are valued in the universal enveloping algebra of the $G$ Lie algebra. 
Due to the SW map, the degrees of freedom of $\Ah$ are the same as that of $A$.
Similarly to $\Ah$, also $\Fh$ is valued in the universal enveloping 
algebra of $G$, and we write
\eq
\Fh_{\mu\nu}=\sum_{s=1}^\infty \,\sum_{\,\,a_1,\ldots,a_s}
{{\cal F}_{\!\mu\nu}^{(a_1,\ldots,a_s)}(\th,\partial,A^{(s)})}\;
T^{a_1}T^{a_2}\ldots T^{a_s}~~\label{FUEA}
\en
where ${{\cal F}_{\!\mu\nu}^{(a_1,\ldots,a_s)}(\th,\partial,A^{(s)})}$
is a function homogeneous and of order $s$ in the gauge potentials
$A_\mu^a$. By dimensional analysis it is at least of order $s$ in $\th$.
{}From (\ref{FUEA}) it is clear that
expression (\ref{Action1multiplet}) is ambiguous because 
in  $T^{\!}r(\Fh_{\mu\nu}\*\Fh^{\mu\nu})$ we have not specified 
the representation $\rho(T^a)$\footnote{We denote with $T$ an
 abstract Lie algebra generator, with $\rho(T)$ a generic representation
 normalized such that $\Tr(\rho(T^a)\rho(T^b))={1\over 2}\delta^{ab}$, 
and with $t$
  a generator in the fundamental representation.}. 
We can render explicit the ambiguity in (\ref{Action1multiplet}) 
by writing
\eq
{1\over g^2}T^{\!}r(\Fh_{\mu\nu}\*\Fh^{\mu\nu})=
\sum_\rho c_\rho\Tr(\rho(\Fh_{\mu\nu})\*\rho(\Fh^{\mu\nu}))
\label{ambiguity}
\en
where the sum is extended over all unitary irreducible and
inequivalent representations $\rho$ of $G$. 
The real coefficients $c_\rho$ parametrize the ambiguity in 
(\ref{ambiguity}). They are constrained by the condition
$$
{{1\over g^2}}=\sum_\rho c_\rho \Tr(\rho(T^a)\rho(T^a))
$$
that is obtained by requiring that in the commutative limit, 
$\th\rightarrow 0$, (\ref{ambiguity}) becomes
the usual commutative gauge kinetic term 
${1\over 2g^2} \sum_a F^a_{\mu\nu}F^{a\,\mu\nu}$.
Notice that in Euclidean space the action (\ref{Action1multiplet})
should be negative definite. Now for each irrep. we have 
\eq
\int d^4x\, 
\Tr(\rho(\Fh_{\mu\nu})\*\rho(\Fh^{\mu\nu})) =
\int d^4x\, 
\Tr(\rho(\Fh_{\mu\nu})\rho(\Fh^{\mu\nu})) =
\int d^4x\, 
\Tr(\rho(\Fh_{\mu\nu})^{\,\dagger}\rho(\Fh^{\mu\nu}))\geq 0
\en
because $\rho(\Fh_{\mu\nu}){}^{\,\dagger}\rho(\Fh^{\mu\nu})$ 
is an hermitian positive operator. In particular  
(\ref{Action1multiplet}) is negative definite if 
the coefficients $c_\rho$ are positive.
\sk
The ambiguity (\ref{ambiguity}) in the action (\ref{Action1multiplet})
can also be studied by expanding (\ref{ambiguity}) in terms of the
commutative fields $\Psi,A,F$. By using (\ref{SWA}),(\ref{SWPsi}) one
obtains (\cite{Jurco:2001rq,SM}),
$$
\widehat{S}_{gauge}=
-{1\over 2 g^2}^{\,}\int d^4x \, 
T^{\!}r(\Fh_{\mu\nu}\*\Fh^{\mu\nu})\nonumber  
~~~~~~~~~~~~~~~~~~~~~~~~~~~~~~~~~~~~~~~~~~~~~~~~~~~~~~~~~~~~~~~~~~~
$$
$$
=-\frac{1}{4g^2}^{\,}\int \!d^4x 
\sum_{a=1}^{{\mbox{{\tiny{dim }}}}{\!G}}
F^a_{\mu \nu}F^{a\,\mu \nu}_{\,}
+{\theta^{\mu \nu} \over 4g^2} \! \int \!d^4x^{\,}   T^{\!}r (F_{\mu \nu}
F_{\rho \sigma} F^{\rho \sigma})^{\,} -   {\theta^{\mu\nu}\over g^2} 
\int \!d^4x \, T^{\!}r (F_{\mu \rho} F_{\nu \sigma}
F^{\rho\sigma}) \, +{\mathcal O}(\theta^2)
$$
The cubic terms in this formula can be further simplified by observing that
\eq
\th^{\mu\nu}F^a_{\mu\rho}F^{b\,}_{\nu\sigma}F^{c\,\rho\sigma}
\,T^{\!}r(T^a[T^b,T^c])=0\label{symmetric}
\en
(use $T^{\!}r(T^a[T^b,T^c])=T^{\!}r(T^c[T^a,T^b])$ and
that $\th^{\mu\nu}F^a_{\mu\rho}F^b_{\nu\sigma}F^{c\,
  \rho\sigma}$ is symmetric in $a\longleftrightarrow b$).
We thus arrive at the expression 
$$
\widehat{S}_{gauge}+{\mathcal O}(\theta^2)=
~~~~~~~~~~~~~~~~~~~~~~~~~~~~~~~~~~~~~~~~~~~~~~~~~~~~
~~~~~~~~~~~~~~~~~~~~~~~~~~~~~~~~~~~~~~~~~~~~~~~~~~~~~~~
~~~~~~~~~~~~~~~~~~~~~~~~~~~~~~~~~~~~~~~~~~~~~~~~~~~~~~~\nonumber\\
$$
$$
=-\frac{1}{4g^2}\int \! d^4x 
\sum_{a=1}^{{\mbox{{\tiny{dim }}}}{\!G}}
F^a_{\mu \nu}F^{a\,\mu \nu}_{\!}
+ {\theta^{\mu \nu}\over 8g^2}\!\int \!d^4x^{\,} T^{\!}r (F_{\mu \nu}
\{F_{\rho \sigma}, F^{\rho \sigma}\})^{\,} - \,{\theta^{\mu\nu}\over 2g^2}\!
\int \!d^4x \, T^{\!}r (F_{\mu \rho} \{F_{\nu \sigma},
F^{\rho\sigma}\})
$$
\eq
{\!\!\!\!}=-\frac{1}{4g^2}\int \! d^4x 
\sum_{a=1}^{{\mbox{{\tiny{dim }}}}{\!G}}
F^a_{\mu \nu}F^{a\,\mu \nu}_{\!}
\,+\,(_{\,}{\mbox{$\sum_\rho$} c_\rho D_\rho^{abc}}) \,\,\,
{\theta^{\mu \nu}\over 4}\!\int \!d^4x^{\,\,\,} {1\over 4}F^a_{\mu \nu}
F^b_{\rho \sigma} F^{c\,\rho \sigma}\, -\,F^a_{\mu \rho} F^b_{\nu \sigma}
F^{c\,\rho\sigma}~~~~~~~~~
\label{traccia}
\en
where 
\eq
{1\over 2}D_\rho^{abc}\equiv\Tr(\rho(T^a)\{\rho(T^b),\rho(T^c)\})
= {\cal A(\rho)}\Tr(t^a\{t^b,t^c\})\equiv
{1\over 2}{\cal A(\rho)}d^{abc}\label{Dterm}~.
\en
Here $t^a$ denotes the fundamental representation, and we are
using that the completely symmetric $D_\rho^{abc}$ tensor in the
representation $\rho$ is proportional to the $d^{abc}$ one defined
by the fundamental representation. In particular 
for all simple Lie groups, except $SU(N)$ with $N\geq 3$, 
we have $D_\rho^{abc}=0$
for any representation $\rho$.
Thus from  (\ref{traccia}) we see that at first order in $\th$  
the ambiguity (\ref{ambiguity}) is present just for  $SU(N)$ Lie
groups, and it is equivalent to the choice
of the real number $\sum_\rho c_\rho{\cal A(\rho)}$. 
\sk
Among the possible representations that one can choose in 
(\ref{ambiguity}) there are two natural ones. 
The fermion representation and the adjoint representation. 

The adjoint representation is particularly appealing if we just have a pure
gauge action, then, since only the structure constants appear in the
commutative gauge kinetic term $\sum_a F_{\mu\nu}^aF^{a\,\mu\nu}$, 
a possible choice is indeed to consider only the
adjoint representation. This is a minimal choice in the sense that in
this case only structure constants enter (\ref{FUEA}) and (\ref{ambiguity}).
In Subsection 3.1 we show that in this case the gauge action is even
in $\th$.

If we also have matter fields then from (\ref{covder})
we see that we must consider the particle representation  
$\rho_\Psi$ given by the multiplet $\Psi$ (and inherited by $\Psih$). 
In (\ref{ambiguity}) one could then make the minimal choice
of selecting just the $\rho_\Psi$ representation.
\sk
Along the lines of the above NCYM theories framework we now examine
the $SO(10)$, the $SU(5)$  and the Standard Model 
noncommutative gauge theories.
\sk
\sk
\noi{\bf 2.1$~~$Noncommutative \mbox{\boldmath ${ SO(10)}$}}
\sk
\noi We first consider only one fermion generation because 
this fits in one multiplet: the $16$-dimensional spinor representation of
$SO(10)$ usually denoted $16^+$.
We write the left handed multiplet as 
\eq
\Psi_L^+=(u^i,d^i\,,\,-u^{{\scriptscriptstyle C}}_i\,,\,
d^{{\scriptscriptstyle   C}}_i\,,\,\nu,e^-\,,\,e^+\,,\,
-\nu^{\scriptscriptstyle  C})_L
\label{ml}
\en
where $i$ is the $SU(3)$ color index and 
$\nu^{\scriptscriptstyle  C}_{~L}=
-i\sigma_2{_{_{\,}}}{\nu_{_{\!}R}}^{^{_{_{_{\scriptstyle *}}}}}$
is the charge conjugate of the neutrino particle $\nu_R$ 
(not present in the Standard Model).
The gauge and fermion sector of noncommutative $SO(10)$ is then simply
obtained by replacing  $\Psih$ with $\Psih^+_L$ in (\ref{Action1multiplet}).

Next we consider all three fermion families. In this case we do not
have a single multiplet, the kinetic term for the fermions reads
\eq
\sum_{B=1}^3
\overline{{{\widehat{ {\Psi\!}^+_L } }}
{^{{}^{^{_{_{_{_{(B)}}}}}}}}} \star i
\widehat{\fmslash D}_{_{\,}} { {\widehat{ {\Psi\!}^+_L }} 
{^{{}^{_{{{(B)}}}}}}}\label{kft}
\en
and in principle in the gauge kinetic term we could have  different
weights $c_{\rho_{\Psi^{(B)}}}$ for each of the three $16$-dimensional 
representations $\rho_{\Psi^{(1)}},\rho_{\Psi^{(2)}},\rho_{\Psi^{(3)}}$.
But these representations are three identical copies and therefore
only the combination  
$c_{\rho_{\Psi^{(1)}}}+c_{\rho_{\Psi^{(2)}}}+c_{\rho_{\Psi^{(3)}}}$
enters.

In conclusion we expect the noncommutative $SO(10)$ gauge kinetic term to
contain at most a combination of the adjoint representation and of the 
$16^+$ particle representation; indeed it is difficult to conceive a 
mechanism that generates other representations than these two. This
holds expecially if one considers the noncommutative $SO(10)$ action
as a fundamental one in the sense that no other fermion has been 
integrated out in order to obtain (\ref{Action1multiplet}).

Finally let us repeat that in (\ref{traccia}),  
whatever representation $\rho$ one considers,
no linear term in $\th$, i.e. no cubic term in $F$ can appear. 
This is so because $SO(10)$ is anomaly free: $D_\rho^{abc}=0$ forall $\rho$.
In other words, at first order in $\th$, noncommutative $SO(10)$ gauge
theory is unique. 
\sk
\sk
\noi{\bf 2.2$~~$Noncommutative \mbox{\boldmath ${ SU(5)}$}}
\sk
\noi The fermionic sector of $SU(5)$ is made by two multiplets 
for each family. 
The ${\psi^{\scriptscriptstyle C}}_L$ multiplet 
transforms in the $\overline{5}$ of $SU(5)$, while the 
$\chi_L$ multiplet 
transforms according to the $10$ of $SU(5)$.

In this case we expect that the adjoint, the ${\overline 5}$
and the $10$ representations enter in (\ref{ambiguity}). 
In principle one can consider the coefficients
$c_{\overline 5}\not=c_{10}$, i.e. while the 
$({\psi^{\scriptscriptstyle C}}_L,\chi_L)$ 
fermion rep.
is $\overline 5\oplus 10$, in (\ref{ambiguity}) the weights
$c_\rho$ of the  ${\overline 5}$ and the $10$ can possibly be not the same.
Only if $c_{\overline 5}\not=c_{10}$ then  
$\sum_\rho c_\rho D_\rho^{abc}\not=0$ 
in (\ref{traccia}). {\sl Proof}: 
The adjoint rep. $C^a$, defined by $[t^a,t^b]=C^{a\,bc}t^c$, is 
antisymmetric, $(C^a)^t=-C^a$, and  therefore we have
\eq
\Tr(C^a\{C^b,C^c\})=\Tr(C^a\{C^b,C^c\})^t=-\Tr(C^a\{C^b,C^c\})=0.
\label{adjoint}
\en 
The representation ${\overline 5}\oplus 10$ is 
anomaly free because $D^{abc}_\5bar=-D^{abc}_{10}$. 
If we consider $c_{\overline 5}\not=c_{10}$ then 
$\sum_\rho c_\rho D_\rho^{abc}=(c_{\overline 5}-c_{10})
D^{abc}_\5bar\not=0$.

We see that, already at first order in the noncommutativity 
parameter $\th$,  noncommutative $SU(5)$ gauge theory 
is not uniquely determined by the gauge coupling constant $g$, 
but also by the value of $\sum_\rho c_\rho D_\rho^{abc}$.
It is tempting to set $c_{\overline 5}=c_{10}$ so that 
$\sum_\rho c_\rho D_\rho^{abc}=0$ and exactly the
fermion representation ${\overline 5}\oplus 10$ (and eventually the
adjoint one) enter (\ref{traccia})\footnote{In \cite{Deshpande} only
$c_{\overline 5}\not=0$ and therefore triple gauge boson couplings 
like $\th FFF$ are present.}. (But this relation is not protected 
by symmetries.) In conclusion we see that $SU(5)$ is \emph{not} a truly unified
theory in a noncommuative setting.
\sk 
\sk
\noi {\bf 2.3$~~$(GUT inspired) Noncommutative Standard Model}
\sk
\noi 
In this case the group is not a simple group.
We denote by $\Tcal^A$ the generators of 
$U(1)\otimes SU(2)\otimes SU(3)$.
They are  $\{\Tcal^A\}=\{ Y,T^a_L,T^\ell_S\}$ with $a=2,3,4$ and $
\ell=5,\ldots, 12$.
Any irrep. of $ U(1)\otimes SU(2)\otimes SU(3)$, is a product of an irrep.
of  $U(1)$, of  $SU(2)$ and of $SU(3)$. We write
$$\{\rho(\Tcal)^A\}=\{\rho_1(Y)\otimes\1_{\rho_2}\otimes\1_{\rho_3}\, ,\,
1\otimes\rho_2(T^a_L)\otimes\1_{\rho_3}\,,\,1\otimes\1_{\rho_2}
\otimes\rho_3(T^b_S)\}.$$
We also have
\eq
\Tr(\rho(\Tcal^1)\rho(\Tcal^a))=\Tr(\rho(\Tcal^1)\rho(\Tcal^\ell))=
\Tr(\rho(\Tcal^a)\rho(\Tcal^\ell))=0 \label{trTT}
\en
because 
$\Tr\left((\rho_1(Y)\otimes\1_{\rho_2}\otimes\1_{\rho_3})
(1\otimes\rho_2(T^a_L)\otimes\1_{\rho_3})\right)=
\Tr(\rho_1(Y))\Tr(\rho_2(T^a_L))\Tr(\1_{\rho_3})=0$
and similarly for $\Tr(\rho(\Tcal^1)\rho(\Tcal^\ell))$ and 
$\Tr(\rho(\Tcal^a)\rho(\Tcal^\ell))$.
The gauge kinetic term is
\eqa
\widehat{S}_{gauge}&=&
-{1\over 2}\int\! d^4x\;\sum_\rho c_\rho\Tr(\rho(\Fh_{\mu\nu})\*\rho(\Fh^{\mu\nu}))\label{SMgauge}\\
&=&-{1\over 2}\int\! d^4x\;\sum_{\rho_1,\rho_2,\rho_3} \!\!c_{\rho_1\otimes\rho_2\otimes\rho_3}
\Tr({(\rho_1\otimes\rho_2\otimes\rho_3})(\Fh_{\mu\nu})\,\*\,
{(\rho_1\otimes\rho_2\otimes\rho_3})(\Fh^{\mu\nu}))
\nonumber
\ena
where the sum is over all inequivalent irrep.'s of 
$U(1)\otimes SU(2)\otimes SU(3)$. 
In the commutative limit only terms quadratic in ${\cal T}$ enter 
the traces, and  using (\ref{trTT}) we obtain
\eqa
\widehat{S}_{gauge}\,{\stackrel{\theta\rightarrow 0~}{-\!\!\!\!\longrightarrow}}\;\,S^{cl}_{gauge}
&=&-{1\over 2}\int\! d^4x\,~\left[{}^{^{^{}}}_{{}_{_{}}}\right. 
\!\sum_{\rho_1,\rho_2,\rho_3}
c_{\rho_1\otimes\rho_2\otimes\rho_3}
\dim (\rho_2) \;\dim (\rho_3)
\;\rho_1(Y)\rho_1(Y)\!_{\!}\left.{}^{^{^{}}}_{{}_{_{}}}
\right]\, F^1_{\mu\nu} F^{1\,\mu\nu}\nonumber\\
&&
{}~~~~~~~~~~~~~+{1\over 2}\left[{}^{^{^{}}}_{{}_{_{}}}\right.\! \sum_{\rho_1,\rho_2,\rho_3}
\!c_{\rho_1\otimes\rho_2\otimes\rho_3}
\dim (\rho_3)\;
\!_{\!}\left.{}^{^{^{}}}_{{}_{_{}}}\right]
\sum_aF^a_{\mu\nu}F^{a\,\mu\nu}\nonumber\\
&&
{}~~~~~~~~~~~~~+{1\over 2}\left[{}^{^{^{}}}_{{}_{_{}}}\right. \!\sum_{\rho_1,\rho_2,\rho_3}
\!c_{\rho_1\otimes\rho_2\otimes\rho_3}
\dim (\rho_2)
\left.{}^{^{^{}}}_{{}_{_{}}}\right]
 \sum_\ell F^\ell_{\mu\nu}F^{\ell\,\mu\nu}\nonumber\\
&\equiv&- \int\! d^4x\;{1\over 4g'^2}F^1_{\mu\nu} F^{1\,\mu\nu}
+{1\over 4g^2}\sum_a F^a_{\mu\nu} F^{a\,\mu\nu}
+{1\over 4g_S^2}\sum_m F^\ell_{\mu\nu} F^{\ell\,\mu\nu}{~\;}
\label{climit}
\ena
where $\dim(\rho)$ is the dimension of the irrep. $\rho$. 
{}From the last line we read  off the 
three  structure constants $g'^2,g^2,g^2_S$
in terms of the coefficients $c_{\rho_1\otimes\rho_2\otimes\rho_3}$
and of the irrep. $\rho_1,\rho_2,\rho_3$.\footnote{With
these general formulas one can recover the results of \cite[Appendix C]{SM}.}
\sk
At first order in $\th$ the only nonvanishing symmetric traces are
$\Tr(\rho({\cal T}^1)\rho(\Tcal^a)\rho(\Tcal^{a'}))$,
$\Tr(\rho({\cal T}^1)\rho(\Tcal^\ell)\rho(\Tcal^{\ell'}))$,
$\Tr(\rho({\cal T}^\ell)\{\rho(\Tcal^{\ell'}),\rho(\Tcal^{\ell''})\})$.
We now recall (\ref{SMgauge}) and (\ref{traccia}) and obtain
\eqa
\widehat{S}_{gauge}=S^{cl}_{gauge}&+&
\!\!\int \!d^4x~\;\nu_1 (\th\, \pu F^1\,F^1\pu F^1 \,+\,
\th\,\pu \tilde{F}^1\,F^1\pu \tilde{F}^1)\nonumber\\
&&+^{\,}\nu_2 \!\sum_a(\th\,\pu F^1\,F^a\pu F^a \,+\,
2\th\,\pu F^a\,F^1\pu F^a \,+\,
\th\,\pu \tilde{F}^1\,F^a\pu \tilde{F}^a\,+\,
2\th\,\pu \tilde{F}^a\,F^1\pu \tilde{F}^a)\nonumber\\
&&+^{\,}\nu_3\! \sum_\ell(\th\,\pu F^1\,F^\ell\pu F^\ell\,+\,
2\th\,\pu F^\ell\,F^1\pu F^\ell \,+\,
\th\,\pu \tilde{F}^1\,F^\ell\pu \tilde{F}^\ell\,+\,
2\th\,\pu \tilde{F}^\ell\,F^1\pu \tilde{F}^\ell)\nonumber\\
&&+{1\over 4}\!\!\!\sum_{\;\rho_1,\rho_2,\rho_3}\!\!\!\!c_{\rho_1\otimes
\rho_2\otimes\rho_3}\dim(\rho_2)D^{\ell^{{}^{\,{^{\!}}}} \ell' \ell''}_{\rho_3}{}^{\!}{\theta^{\mu \nu}}[ {1\over 4}F^\ell_{\mu \nu}
F^{\ell'}_{\rho \sigma} F^{{\ell''}\rho \sigma}\, -\,F^\ell_{\mu
  \rho} 
F^{\ell'}_{\nu \sigma}
F^{{\ell''}\rho\sigma}]\label{firstorderSM}
\ena
where $\th\,\pu F\equiv \th_{\mu\nu}F^{\mu\nu}$,
$\th\,\pu \tilde{F}\equiv \th_{\mu\nu}\tilde{F}^{\mu\nu}$ and
$\tilde{F}^{\mu\nu}\equiv{1\over 2}\, \varepsilon^{\mu\nu\rho\sigma}F_{\rho\sigma}$. 
In (\ref{firstorderSM}) we have used 
\eqa
\th_{\nu\mu}F^{1\,\mu\rho}F^a_{\rho\sigma}F^{a\,\sigma\nu}
&=&{1\over 4}[\th\,\pu F^1\;F^a\pu F^a\,+\,\th\,\pu F^a\;F^1\pu F^a
\,+\,\th\,\pu \tilde{F}^a\;F^1\pu \tilde{F}^a]\nonumber\\
\th_{\nu\mu}F^{a\,\mu\rho}F^1_{\rho\sigma}F^{a\,\sigma\nu}
&=&{1\over 2} \th\,\pu F^a\;F^1\pu F^a\,
+\,{1\over 4}\th\,\pu \tilde{F}^1\;F^a\pu \tilde{F}^a
\nonumber
\ena
and similar formulas  with $a\rightarrow \ell$ and  $a\rightarrow 1$. 
The coefficients $\nu_1$,$\nu_2$,$\nu_3$ 
depend on $c_{\rho_1\otimes\rho_2\otimes\rho_3}$
and the  irrep. $\rho_1,\rho_2,\rho_3$.

In the particle representations, only the trivial, the fundamantal 
and the conjugate of the fundamental appear for  colour $SU(3)$.
(This accounts for invariance under charge conjugation: 
if we replace $u$ and  $d$ with $u^{\scriptscriptstyle C}$ and 
$d^{\scriptscriptstyle C}$ the $SU(3)$ 
lagrangian is unaffected).
In the noncommutative case it is natural to preserve this
symmetry between representations and conjugate representations. 
In other words any irrep. $\rho_3$ of  $SU(3)$ should appear together 
with its conjugate irrep. $\rho^*_3$ and with
$c_{\rho_1\otimes\rho_2\otimes\rho^*_3}
=c_{\rho_1\otimes\rho_2\otimes\rho_3}$. The last line in 
(\ref{firstorderSM}) then vanishes. The  proof easily follows
from 
$$
{1\over 2}D_{\rho^*_3}^{\ell^{{}^{\,{^{\!}}}} \ell' \ell''}
=\Tr(\rho^*_3(\Tcal^\ell) \{\rho^*_3(\Tcal^{\ell'}),
\rho^*_3(\Tcal^{\ell''})\})=-
\Tr(\rho_3(\Tcal^\ell) \{\rho_3(\Tcal^{\ell'}),
\rho_3(\Tcal^{\ell''})\})=-{1\over 2}D_{\rho_3}^{\ell^{{}^{\,{^{\!}}}}
  \ell' \ell''}
$$
where we used $\rho^*(\Tcal)=-\rho(\Tcal)$ since $\Tcal$ is hermitian. 
\sk
About the fermion kinetic term, the fermion vector  
${\widehat \Psi_L^{(B)}}$, where $B=1,2,3$ 
is the family index, is given by (cf. $\Psi_L^+$ above)
$\Psi_L=(u^i,d^i\,,\,-u^{\scriptscriptstyle C}_i\,,
\,d^{\scriptscriptstyle C}_i\,,\,\nu,e^-\,,\,e^+)_L$.
The  covariant derivative is as in (\ref{covder}), 
with $\Psi\rightarrow \Psi_L$ and with $A_\mu=A_\mu^A\Tcal^A$.
The fermion kinetic term is then as in (\ref{kft}) 
(with $\Psi_L^+\rightarrow \Psi_L$).
This Standard Model is built using only left handed fermions and
antifermions. We call it GUT inspired because its noncommutative
structure can be embedded in $SO(10)$ GUT. Indeed $\Psi_L$ and 
$\Psi_L^+$  differ just by the extra neutrino 
$\nu^{\scriptscriptstyle  C}_{~L}=
-i\sigma_2{_{_{\,}}}{\nu_{_{\!}R}}^{^{_{_{_{\scriptstyle *}}}}}$;
moreover under an infinitesimal gauge transformation $\lh$,
all fermions in $\Psi_L$ transform with $\lh$ on the left.
This GUT inspired Standard Model differs from the one considered 
in \cite{SM}; indeed here we started from the chiral vector
$\Psi_L$, while there the vector $\Psi'=(u^i_L,d^i_L\,,\,u^i_R\,d^i_R\,,\,\nu_L,e^-_L\,,\,e^-_R)$
is considered. In the commutative case $\int \overline{\Psi'} 
{\fmslash D} \Psi'=\int \overline{\Psi_L}  
{\fmslash D} \Psi_L$ but in the noncommutative case, as we discuss in 
Subection 4.1 (see also (\ref{er}) and the last lines of
the next subsection), this is no more true:  
$\int \overline{\widehat{\Psi'\,}} \*
{\widehat{\fmslash D}} \widehat{\Psi'\,}\not=
\int \overline{\widehat{\Psi_{\!L}}} \*
{\widehat{\fmslash D}}_{_{\,}} \widehat{\Psi_{\!L}}\,$,
if we change  $\th$ into $-\th$  in the right handed sector of 
$\int \overline{\widehat{\Psi'\,}} \*
{\widehat{\fmslash D}} \widehat{\Psi'\,}$, 
then the two expressions coincide.
\sk
\sk
Finally, if in the gauge kinetic term (\ref{firstorderSM})
we consider only the adjoint rep. and the fermion rep.
we have  
\eq
\widehat{S}_{gauge}=S^{cl}_{gauge}+{\mathcal O}(\th^2).
\en
This is so because the fermion rep. is anomaly free:
$D_{\rho_{{_{\rm{fermion}}}}}^{A^{\,_{\!\!}}A'\!A''}=0$ 
($A=1,a,l$), because for $U(1)$ the adjoint rep. is trivial, 
and, for the last line in (\ref{firstorderSM}), 
because the adjoint rep. of $SU(3)$ has also
$D_{{\rho_3}_{\rm{adj.}}}^{\ell^{{}^{\,{^{\!}}}}\ell' \ell''}=0$ 
(a proof is as in (\ref{adjoint})).

\sk 
\sk
\sk
\noi {\bf 2.4$~~$Higgs Sector in the (GUT inspired) Noncommutative Standard Model}
\sk
\noi 
In the commutative SM, the Yukawa terms can be written as 
\eq
W{^{{}^{_{{{BB'}}}}}}\phi^\dagger {L_L}{^{{}^{_{{{\!(B)}}}}}}
{e_R^*}{^{{}^{_{{{\!(B')}}}}}}
+
G_u{^{{}^{_{{{\!\!\!BB'}}}}}}\widetilde{\phi}^{\,\dagger} {Q_L}{^{{}^{_{{{\!(B)}}}}}}
{u_R^*}{^{{}^{_{{{\!(B')}}}}}}
+
G_d{^{{}^{_{{{\!\!\!BB'}}}}}}\widetilde{\phi}^{\,\dagger} {Q_L}{^{{}^{_{{{\!(B)}}}}}}
{d_R^*}{^{{}^{_{{{\!(B')}}}}}}
+~herm. ~con_{\!}j.
\label{25}
\en
where 
$L_L^{}=\left({^{^{^{\!}}}}{_{_{_{\!\!}}}}\right.
{}^{^{\mbox{\sma{$\nu^{}_L $}}}}_{_{\mbox{\sma{$e^{}_L $}}}}
\left.{^{^{^{\!}}}}{_{_{_{\!\!\!}}}}\right)\,$,
$Q^{}_L=\left({^{^{^{\!}}}}{_{_{_{\!\!}}}}\right.
{}^{^{\mbox{\sma{$u_L^{} $}}}}_{_{\mbox{\sma{$d^{}_L$}}}}
\left.{^{^{^{\!}}}}{_{_{_{\!\!\!}}}}\right)\,$, 
$\phi^{}=\left({^{^{^{\!}}}}{_{_{_{\!\!}}}}\right.
{}^{^{\mbox{\sma{$\varphi_+ $}}}}_{_{\mbox{\sma{$\varphi_0 $}}}}
\left.{^{^{^{\!}}}}{_{_{_{\!\!\!}}}}\right)\,$
and
$\widetilde{\phi}= i\tau_2\phi^*$ with
$\tau_2$ the Pauli matrix in $SU(2)$ gauge group space.
The matrices $W{^{{}^{_{{{BB'}}}}}},G_u{ ^{{}^{_{{{\!\!\!BB'}}}}}},
G_d{^{{}^{_{{{\!\!\!BB'}}}}}}$ are the Yukawa couplings 
({\small{$B,B'=1,2,3$}}),
and 
the sum over group and spinor indices is understood,
so that $\phi^\dagger {L_L^{}}{e_R^*}
={e_{R\,}^\dagger}\phi^\dagger L_{L}^{}\,$, 
$\,\widetilde{\phi}^\dagger {Q_L^{}}{u_R^*}
={u_{R\,}^\dagger}\widetilde{\phi}^\dagger Q_{L}^{}\,,\,$ etc.. 
\sk
A noncommutative version of (\ref{25}) is not straighforward;
for example, the noncommutative fields 
$$\widehat{\phi}^{\,\dagger}~~,~~~~ 
\widehat{L_L^{}}~~,~~~~\widehat{{e_{R\,}^*}}=\widehat{{i\sigma_2 e_{~L}^C}}=
i\sigma_2 \widehat{{e_{~L}^C}}~,$$
under an infinitesimal $U(1)\otimes SU(2)\otimes SU(3)$ 
gauge transformation
$\Lambda$,
transform as 
\eq
\delta\,\widehat{\phi}^{\,\dagger}=-i\widehat{\phi}^{\,\dagger}\*\rho_\phi(\lh)~~,~~~
\delta\,  \widehat{L_{L}^{}} =i\rho_{L_L^{}}(\lh)*\widehat{L_{L}^{}}~~,
\en
\eq
{}~~~~\delta\,\widehat{{e_{R\,}^*}}=\delta\, i\sigma_2
\widehat{{e_{~L}^C}}=i\rho_{i\sigma_2 e_{~L}^C}(\lh)\* 
\widehat{{i\sigma_2 e_{~L}^C}}=i\rho_{e_R^*}(\lh)\* 
\widehat{{e_R^*}}
\label{er}
\en
and therefore the term $\int \!d^4x\,\widehat{\phi}^\dagger \*
\widehat{L_L^{}}\*{\widehat{e_R^*}}
=\int \!d^4x\,\widehat{{e_{R\,}^*}}^{{\,}t}\*\widehat{\phi}^\dagger\*
\widehat{L_{L}^{}}$ cannot be gauge invariant because, for example, 
$\rho_{e_R^*}(\lh)$ does not commute with $\widehat{L_L^{}}$.

A solution is to consider hybrid Seiberg-Witten 
maps  ${\widehat{{\quad}}}^{^{_{_{_{\!\!H}}}}}\!$ 
\cite{hep2001} on $L_L^{}$ and $Q_L^{}$.
The noncommutative Yukawa terms are then
\eq
W{^{{}^{_{{{BB'}}}}}}\widehat{\phi}^{{{\,}}\dagger} \*
\widehat{L_L}^{^{_{_{_{\!\!H}}}}}\!{^{{}^{_{{{\!(B)}}}}}}\*
\widehat{e_R^*}{^{{}^{_{{{(B')}}}}}}
+\;
G_u{^{{}^{_{{{\!\!\!BB'}}}}}}\widehat{\widetilde{\phi}\,}{}^{\dagger}\,\* 
\widehat{Q_L}^{^{_{_{_{\!\!\widetilde H}}}}}\!{^{{}^{_{{{\!(B)}}}}}}\*
\widehat{u_R^*}{^{{}^{_{{{(B')}}}}}}
+\;
G_d{^{{}^{_{{{\!\!\!BB'}}}}}}\widehat{\phi}^{\,\dagger} \*
\widehat{Q_L}^{^{_{_{_{\!\!H}}}}}\!{^{{}^{_{{{\!(B)}}}}}}\*
\widehat{d_R^*}{^{{}^{_{{{(B')}}}}}}
+~herm. ~con_{\!}j.
\label{hsw}
\en
with $\widehat{\phi}\equiv\widehat{\phi}[\phi,A,\th]$ and
$\widehat{\widetilde{\phi}\,}\equiv\widehat{\widetilde{\phi}\,}[\widetilde{\phi}_{{_{_{_{_{}}}}}},A,\th]$.  
Under an infinitesimal $U(1)\otimes SU(2)\otimes SU(3)$ 
gauge transformation
$\Lambda$, 
${\widehat{{L_L^{}}}}^{^{_{_{_{\!\!H}}}}}\!$,
$\widehat{Q_L}^{^{_{_{_{\!\!\widetilde H}}}}}\!$, and 
$\widehat{Q_L}^{^{_{_{_{\!\!H}}}}}\!$ transform as
\eqa 
&&\delta\,{\widehat{{L_L^{}}}}^{^{_{_{_{\!\!H}}}}}\!=
i\rho_\phi(\lh)\*{\widehat{{L_L^{}}}}^{^{_{_{_{\!\!H}}}}}\!
-i{\widehat{{L_L^{}}}}^{^{_{_{_{\!\!H}}}}}\!\*
\rho_{e_R^*}(\lh)\label{d29}\\
&&\delta\,{\widehat{{Q_L^{}}}}^{^{_{_{_{\!\!\widetilde H}}}}}\!=
i\rho_{\widetilde{\phi}}(\lh)\*{\widehat{{Q_L^{}}}}^{^{_{_{_{\!\!\widetilde H}}}}}\!
-i{\widehat{{Q_L^{}}}}^{^{_{_{_{\!\!\widetilde H}}}}}\!\*
\rho_{u_R^*}(\lh)\\
&&\delta\,{\widehat{{Q_L^{}}}}^{^{_{_{_{\!\! H}}}}}\!=
i\rho_\phi(\lh)\*{\widehat{{Q_L^{}}}}^{^{_{_{_{\!\! H}}}}}\!
-i{\widehat{{Q_L^{}}}}^{^{_{_{_{\!\!H}}}}}\!\*
\rho_{d_R^*}(\lh)\label{d31}
\ena
We see that in the hybrid SW map $\lh$ appears both on the
left and on the right of the fermions. 
We also see that the representation of $\lh$
is inherited from the Higgs and fermions that respectively sandwich 
${\widehat{{L_L^{}}}}^{^{_{_{_{\!\!H}}}}}\!$,
$\widehat{Q_L}^{^{_{_{_{\!\!\widetilde H}}}}}\!$ and 
$\widehat{Q_L}^{^{_{_{_{\!\!H}}}}}\!\!$.
The Yukawa terms (\ref{hsw}) are thus  invariant under
noncommutative gauge transformations. Of course in the $\th\rightarrow
0$ limit (\ref{d29})-(\ref{d31}) become 
$\delta\,L_L^{}=\rho_{L_L^{}}\!(\Lambda)_{\,}L_L^{}\,,\,
\delta\,Q_L^{}=\rho_{Q_L^{}}\!(\Lambda)_{\,}Q_L^{}\,$.
At first order in $\th$ we have \cite{hep2001,SM}
\eq
\widehat{\,\Psi\,}^{^{_{_{_{\!H}}}}}\! = \Psi + \frac{1}{2}\theta^{\mu\nu} A_\nu
\Big(\pp\mu\Psi -\frac{i}{2} (A_\mu \Psi - \Psi A'_\mu)\Big)
+\frac{1}{2}\theta^{\mu\nu} 
\Big(\pp\mu\Psi -\frac{i}{2} (A_\mu \Psi - \Psi A'_\mu)\Big)A'_\nu
+ \mathcal{O}(\theta^2)\label{HSWPsi}\;~  \label{hybridswmap}
\en
where $A$ carries the representation of the fields on the left of 
$\widehat{\,\Psi\,}^{^{_{_{_{\!H}}}}}$, while 
$A'$ carries the representation of the fields on the right of
$\widehat{\,\Psi\,}^{^{_{_{_{\!H}}}}}$.
The choice of Yukawa terms (\ref{hsw}) differs from those studied in \cite{SM}.
There the hybrid SW map is considered on $\phi$, in particular there
${\widehat{\phi}}^{^{^{{{H}}}}}\!$ is not invariant under $SU(3)$
gauge transformations; here, as in the commutative case, 
$\delta\widehat{\phi}=0$ under $SU(3)$ transformations.\footnote{This
implies, that in \cite{SM} gluons couple directly to the Higgs field,
which is not the case here.}
Another main difference (cf. also Subsection 4.1) is that in \cite{SM} 
$\delta\,\widehat{e_{_{\!}R}^{}}^*=-i\widehat{e^{}_{_{\!}R}}^*\*
\rho_{e^{}_{\!R}}(\lh)$,
i.e. contrary to (\ref{er}), $\lh$ appears on the right and not on
the left of the right handed electron.
\sk
Finally the Higgs kinetic and potential terms are given by 
\eq
(\widehat D_\mu \widehat \phi)^\dagger
\star \widehat D^\mu \widehat \phi           
\,+\, \mu^2 {\widehat{\phi}}^{\,\dagger} \star  \widehat \phi - \lambda\,
\widehat{\phi}^{\,\dagger} \star  \widehat{\phi}
\star
{\widehat \phi}^{\,\dagger} \star {\widehat \phi}\,   ~.
\en


\sk 
\sk
\noi {\bf 2.5$~~$Higgs Sector in Noncommutative
                             \mbox{\boldmath ${ SO(10)}$}}
\sk
\noi 
Up to now we have examined three different noncommutative gauge
theories, $SO(10)$, $SU(5)$ and SM. 
One can also consider the spontaneous symmetry breaking 
$SO(10)~\rightarrow G~\rightarrow U(1)\otimes SU(2)\otimes SU(3)~
\rightarrow U(1)\otimes SU(3)$. 
There are many patterns for SSB depending on the choices of the
Higgses and of the intermediate symmetry group $G$ (e.g. 
$SU(4)\otimes SU(2)_L^{}\otimes SU(2)_R^{}$, $\,SU(5)$ etc.,   
see for example \cite{BU}). 
In general one can construct a noncommutative version of a
given Higgs potential using the SW map and the hybrid SM map.
The noncommutative $SO(10)$ invariant Yukawa terms are built using
similar techniques.
In the commutative case we have the Yukawa term \cite{Strocchi}
\eq
i{}\Phi_{10}^*{\Psi_L^{+}}^t\sigma_2\Psi_L^+  -
i{\Psi_L^{+}}^\dagger\sigma_2{\Psi_L^{+}}^*{}_{\,}\Phi_{10}
\label{h10}
\en
where here transposition is just in the spin indices,
$\Phi_{10}^*{\Psi_L^{+}}^t\sigma_2\Psi_L^+ =
\Phi_{10}^*{\Psi_L^{+}}_{\,\al}\Psi_{L\,\be}^+{\,}\sigma_{2\:\al\be}\;;$
moreover we have suppressed the family indices, so that
the  term ${\Psi_L^{+}}_{\,\al}\Psi_{L\,\be}^+$ stands for
$W^{{{}^{{{{{_{BB'}}}}}}}} 
{\Psi_L^{+}}_{\,\al}{{{}^{^{_{_{_{_{\!\!\!\!\!(B)}}}}}}}} 
\Psi_{L\,\be}^+{{{}^{^{_{_{_{_{\!\!\!\!(B')}}}}}}}}$. 
Similar terms are obtained with the Higgs multiplets $\Phi_{126}$
and $\Phi_{120}$.
Here $10$, $126$ and $120$ are the irrep. 
contained in $16\times 16=10\oplus 126\oplus 120$, 
so that the Yukawa term (\ref{h10}) is an $SO(10)$ singlet.
A noncommutative generalization of (\ref{h10}) is obtained requiring
that the noncommutative version of ${\Psi_L^{+}}^t\sigma_2\Psi_L^+$ 
transforms  as $16\times 16$. This is achieved with the term 
\eq
{\widehat{{\Psi_{L}^{+}}}}_\al^{^{_{_{_{\!\!H}}}}}\!\*
\widehat{\Psi_{L}^{+\,}}_\be{_{_{_{}}}}\sigma_{2\,\al\be}=
{\widehat{{{\Psi_{L}^{+\,}}}_\al\!\otimes \1}}^{^{_{_{_{\!\!\!\!\!\!H}}}}}\!\*
{\widehat{\1\otimes {\Psi_{L}^{+}}}}_\be{_{_{_{_{}}}}}\sigma_{2\,\al\be}
\label{35}
\en
where $\1$ is the $16\times 16$ unit matrix, ${\widehat{\1\otimes
{\Psi_{L}^{+}}}}=\1\otimes\widehat{\Psi_L^+}$ 
is the standard SW map on ${\Psi_L^+}$ and 
${\widehat{{{\Psi_{L}^{+\,}}}\!\otimes \1}}^{^{_{_{_{\!\!\!\!\!\!H}}}}}$
is the hybrid SW map (cf. (\ref{HSWPsi}));
by definition, under an infinitesimal gauge transformation (cf. (\ref{deltapsi}))
$\delta{\widehat{{{\Psi_{L}^{+\,}}}}}=
i\rho_{{\Psi_{L}^{+}}}(\lh)\*{\widehat{{{\Psi_{L}^{+\,}}}}}$
and 
$$
\delta\left({\!\!}^{^{}}_{_{}}\right. 
{\widehat{{{\Psi_{L}^{+\,}}}\otimes \1}}^{^{_{_{_{\!\!\!\!\!\!H}}}}}
\left.{\!\!}^{^{}}_{_{}}\right)=
i\Big( \rho_{\Psi_{\!L}^{\!+}}(\lh)\otimes \1+
\1\otimes \rho_{\Psi_{\!L}^{\!+}}(\lh)\Big)\*
\Big({\widehat{{{\Psi_{L}^{+\,}}}\otimes \1}}^{^{_{_{_{\!\!\!\!\!\!H}}}}}\Big)-
i \Big({\widehat{{{\Psi_{L}^{+\,}}}\otimes\1}}^{^{_{_{_{\!\!\!\!\!\!H}}}}}\Big)
\* \Big(\1\otimes \rho_{\Psi_{\!L}^{\!+}}(\lh)\Big)\,.
$$
We see that in the commutative limit
${\widehat{{{\Psi_{L}^{+\,}}}\!\otimes \1}}^{^{_{_{_{\!\!\!\!H}}}}}$ 
transforms as ${\Psi_{L}^{+\,}}$.
The noncommutative Yukawa term then reads
\eq
i_{\,}\widehat{\Phi}^{^{^{_{_{\scriptstyle{\,*}}}}}}_{10}\*{\widehat{{\Psi_{L}^{+}}}}_\al^{^{_{_{_{\!\!H}}}}}\!\*
\widehat{\Psi_{L}^{+\,}}_{\be}{_{_{_{\,}}}}\sigma_{2\,\al\be}\:
-
\,i_{\,}{\widehat{\Psi_{L}^{+\,}}_\al}^{\,*}\*
{{{\widehat{{\Psi_{L}^{+}}}}_\be^{^{_{_{_{\!\!H}}}}}}}
{}^{^{^{^{_{_{\scriptstyle{*}}}}}}}\!\*
\widehat{\Phi}_{10}{_{_{_{\,}}}}\sigma_{2\,\al\be}
\en
where $\delta\widehat{\Phi}_{10} =
i\rho_{10}(\lh)\*{\widehat{\Phi}_{10}}$. Similarly for
$\widehat{\Phi}_{126}$ and $\widehat{\Phi}_{120}$.
\sk
\sk
\section{Hermiticity and reality of SW map}
In the previous section we used that if $A$ is hermitian
then $\Ah$ is hermitian too. At first order in $\th$ this is
indeed the case, see (\ref{SWA}).
In this section we show, to all orders in $\th$, that $\Ah$ and $\lh$ 
can be chosen
hermitian if $A$ and $\Lambda$ are hermitian. 
More generally we show that SW map commutes with hermitian 
conjugation as well as with complex conjugation. 

Given a (not necessarily unitary) 
rep. $\rho_\Psi$ of $G$ defined by the multiplet 
$\Psi$, we  can always consider the multiplet $\U$ that transforms
according to the inverse hermitian representation $\rho_\U$ given by
$\rho_\U(g)\equiv(\rho_\Psi(g))^{{-1}^{\,\mbox{\scriptsize{$\dagger$}}}}$, 
forall $g\in G$. 
Similarly we consider the multiplet $\Psi^*$ that transforms according 
to the conjugate rep. $\rho_{\Psi^*}(g)\equiv\overline{\rho_\Psi(g)}$.
\footnote{In order to avoid possible confusions with 
$\Psi^\dagger \gamma^0$,
on multiplets we denote complex conjugation with $^{*}$ instead of 
${}^{\overline{{{}^{~\;}}}}\,$. Also, in this section, the multiplet
$\Psi$ is not necessarily a fermion multiplet.} 
Since 
$g=e^{i\Lambda}=e^{i\Lambda^aT^a}$, with $\L^a$ real, and $A=A^aT^a$ 
with $A^a$ real, at the Lie algebra level we have
\eq
\rho_\U (\L)=(\rho_\Psi(\L))^\dagger~,~~\rho_\U(A)=(\rho_\Psi(A))^\dagger
{}~~~,~~~~~~~
\rho_{\P*}(\L)=-\overline{\rho_\Psi(\L)}
{}~,~~\rho_\P*(A)=-\overline{\rho_\Psi(A)}\;.\label{commU}
\en
Commutativity of SW map with hermitian conjugation and with complex
conjugation means
\eq
\widehat{\rho_\U (A)}={\widehat{\rho_\Psi(A)\,}}^\dagger~,~~
\widehat{\rho_\U (\L)}={\widehat{\rho_\Psi(\L)\,}}^\dagger~~
~~~\mbox{i.e.}~~~~~
\widehat{(\rho_\Psi (A))^{{^{{\!}}}\dagger}{}~{}}\!={\widehat{\rho_\Psi(A)\,}}^\dagger~,~~
\widehat{(\rho_\Psi (\L))^{{^{{\!}}}\dagger}{}~{}}\!={\widehat{\rho_\Psi(\L)\,}}^\dagger{}~,
\label{her}
\en
that for short we rewrite $\widehat{A^\dagger}=
\Ah^{^{{{\,{\scriptstyle{\dagger}}}}}}{}\,,~
\widehat{\L^\dagger}=\lh^{^{{{\,{\scriptstyle{\dagger}}}}}}{}\,$,  and
\eq
\widehat{\P*}={\widehat{\Psi}}^{{^{\,\mbox{\scriptsize{$*$}}}}}~~,~~~
\widehat{\rho_\P* (A)}=-\overline{{\widehat{\rho_\Psi(A)\,}}}~~,~~~
\widehat{\rho_\P* (\L)}=-\overline{{\widehat{\rho_\Psi(\L)\,}}}~~.~~
\label{conj}
\en
In (\ref{her}) and (\ref{conj}) 
we used the following notation (cf. Section 2)
\eqa
\widehat{\rho_\Psi (A)}\equiv{\rho_\Psi
  (\Ah)}\equiv\Ah[\rho_\Psi(A),\th]~~~~~~~&,&~~~
\widehat{\rho_\Psi (\L)}\equiv{\rho_\Psi (\lh)}
\equiv\lh[\rho_\Psi(A),\rho_\Psi(\L),\th]\label{27}\\
\widehat{\rho_\U (A)}\equiv{\rho_\U
  (\Ah)}\equiv\Ah[\rho_\U(A),\th]~~~~~~~&,&~~~
\widehat{\rho_\U (\L)}\equiv{\rho_\U (\lh)}
\equiv\lh[\rho_\U(A),\rho_\U(\L),\th]\label{28}\\
\widehat{\rho_\P* (A)}\equiv{\rho_\P*
  (\Ah)}\equiv\Ah[\rho_\P*(A),-\th]~&,&~~~
\widehat{\rho_\P* (\L)}\equiv{\rho_\P* (\lh)}
\equiv\lh[\rho_\P*(A),\rho_\P*(\L),-\th]~~~~~~\label{29}
\ena
\eq
\Psih\equiv \SW [\Psi,\rho_\Psi(A), \th]~~~,~~~~~~
\widehat{\U}\equiv \SW [\U,\rho_\U(A), \th]~~~,~~~~~~
\widehat{\P*}\equiv \SW [\P*,\rho_\P*(A), - \th]
\label{30}
\en
Notice that in (\ref{29}) $\th$ appears with the opposite sign
w.r.t. $\th$ in (\ref{27}), similarly in (\ref{30}).
Consistency requires that with the representation
$\rho_\P*$, we must consider the  opposite star product 
$\m*$ i.e. the star product built with $-\th$ instead of $\th$. The $-\th$ in (\ref{29}) 
is also consistent with the charge conjugation operator  defined
in (\ref{ACC}).

{}For the proof of (\ref{her}) and (\ref{conj}), in the case of
constant $\th$,  we can use SW differential
equation \cite{SW}.  As discussed in  \cite{hep-th/9909139},
SW differential equation is not unique. Hermiticity and reality
indeed constrain the freedom in the choice of SW map. 
Hermiticity and reality are physical
requirements, since we want the noncommutative action 
(obtained via SW map from the commutative one) to be real 
if the commutative one is real.
However there are two extra ambiguities. 
One is related to gauge transformations, SW map is 
defined to map orbits of the commutative gauge group to orbits of the 
noncommutative one, therefore there is no unique way to associate to a
given commutative gauge potential a given noncommutative gauge
potential. The other ambiguity is related to field redefinitions 
of the noncommutative gauge potential. 

We expect  that in both cases
these ambiguities are not physical because the noncommutative 
$S$ matrix is gauge invariant and is expected to be independent 
from field redefinitions.

Here we choose a specific SW differential equation, it reads \cite{SW}:
\eqa
&&\delta_\th\Ah_\mu=\delta\th^{\rho\sigma}
{\partial\over\partial\th^{\rho\sigma}}
\Ah_\mu=-{1\over 4}\delta\th^{\rho\sigma}
\{\Ah_\rho \,{\stackrel{\displaystyle \*}{,}}\, 
\partial_\sigma\Ah_\mu+\Fh_{\sigma\mu}\}
\label{one}\\
&&
\delta_\th\lh=\delta\th^{\rho\sigma}
{\partial\over\partial\th^{\rho\sigma}}\lh
=-{1\over 4}\delta\th^{\rho\sigma}
\{\partial_\rho\lh \,{\stackrel{\displaystyle \*^{}}{,}}\, \Ah_\sigma\}
\label{two}
\ena
in these expressions  $\Ah$ and $\lh$ are valued in 
the universal enveloping algebra of $G$. 
As in \cite{SW}, (\ref{one}) and (\ref{two}) are obtained by requiring
that gauge equivalence classes of  the ${\Ah}^{{'}}$ gauge
theory, with noncommutativity  ${\*{}^{\!}}'$ given by 
$\th'\equiv\th+\de_\th\th$, correspond to 
gauge equivalence classes of  the ${\Ah}$ gauge
theory, with noncommutativity ${\*}$ given by 
$\th$. In formulas, writing ${\Ah}^{{'}}={\Ah}^{{'}}[A,\th']
={\Ah}^{{'}}[\Ah,\de_\th\th]$ the condition reads
\eq
\de_{{\lh}^{{\!'}}}{\Ah}^{{'}}
\equiv \partial\lh^{{{_{_{_{\!\!}}}}'}}+\,i[{\lh}^{{{_{_{_{\!\!}}}}'}}
\,{\stackrel{\displaystyle \*^{{\!}}{^{_{_{\scriptstyle '\!\!}}}}}{,}}
\,{\Ah}^{{'}}]
=\de_\lh{\Ah}^{{'}}[\Ah,\de_\th\th]\,\,,
\en
where $\de_\lh{\Ah}^{{'}}[\Ah,\de_\th\th]\equiv 
{\Ah}^{{'}}[\Ah +\de_\lh{\Ah},\de_\th\th]_{_{_{{_{_{_{}}}}}}}-{\Ah}^{{'}}[\Ah,\de_\th\th]$.
SW map for the multiplet  $\Psi$ can similarly be obtained 
by requiring 
$\de_{{\lh}^{{'}}}{\Psih}^{\!{'}}
\equiv\,i{\lh}^{{'}}\,{\*{}^{\!}}'\,{\Psih}^{\!{'}}
=\de_\lh{\Psih}^{\!{'}}[\Ah,\Psih]\,$;  we  have
\eq
\delta_\th\Psih = \delta\th^{\rho\sigma}
{\partial\over\partial\th^{\rho\sigma}}\Psih
=
-\frac{1}{2} \delta\theta^{\mu\nu}\widehat{\rho_\Psi(A_\mu)}\*\pp\nu\Psih
+\frac{i}{8}\delta\theta^{\mu\nu}\s[\widehat{\rho_\Psi(A_\mu)} , 
\widehat{\rho_\Psi(A_\nu)}]\*
\Psih ~.
\label{three}
\en
\sk

In order to show (\ref{her}) we notice that for generic space-time 
dependent matrices $M$ and $N$, under complex conjugation, 
transposition and hermitian conjugation we have
(recall $f\*g=f {_{\,}}e^{{i\over 2}{\th}^{\mu\nu}{\stackrel{\leftarrow}
{\partial}_{\!\mu}}{\stackrel{\rightarrow}{\partial}_{\!\nu}}} {_{\,}}g$)
\eq
\overline{(M\*N)}=\overline{M}\m*\overline{N}
{}~~~~,~~~~~~~(M\*N)^t=N^t\m*M^t
{}~~~~,~~~~~~~(M\*N)^\dagger=N^\dagger\*M^\dagger~~.
\label{mn}
\en
We now apply ${}^\dagger$ 
to (\ref{one}) and (\ref{two}) in the $\rho_\Psi$ representation and
obtain
\eqa
\delta_\th\widehat{\rho_\Psi(A_\mu)}^\dagger
&=&-{1\over 4}\delta\th^{\rho\sigma}
\{{\widehat{\rho_\Psi(A_\rho)}^\dagger \,{\stackrel{\displaystyle\*}{,}}\, 
\partial_\sigma\widehat{\rho_\Psi(A_\mu)}^\dagger+
\widehat{\rho_\Psi(F_{\sigma\mu})}}^\dagger\}
\label{31}\\
\delta_\th\widehat{\rho_\Psi(\L)}^\dagger
&=&-{1\over 4}\delta\th^{\rho\sigma}
\{\partial_\rho{\widehat{\rho_\Psi(\L)}}^\dagger 
\,{\stackrel{\displaystyle\*}{,}}\, 
\widehat{\rho_\Psi(A_\sigma)}^\dagger\}
\label{32}
\ena
If at order ${\mathcal O}(\delta\th^n)$ we have
$\widehat{(\rho_\Psi (A))^{\dagger}{}~}\!\!
={\widehat{\rho_\Psi(A)\,}}^\dagger\,$, 
$\,\widehat{(\rho_\Psi (\L))^{\dagger}{}~}\!\!
={\widehat{\rho_\Psi(\L)\,}}^\dagger\,$, 
then (\ref{31}), (\ref{32}) show that this is also true
for $\widehat{\rho_\Psi(A)}^{\!{_{{{\mbox{\normalsize{$'$}}}}}}}=\widehat{\rho_\Psi(A)}+
\delta_\th\widehat{\rho_\Psi(A)}$ and 
 $\widehat{\rho_\Psi(\L)}^{\,{_{{{\!\!\mbox{\normalsize{$'$}}}}}}}=\widehat{\rho_\Psi(\L)}+
\delta_\th\widehat{\rho_\Psi(\L)}\,$, 
i.e. (\ref{her}) holds also  at order  
${\mathcal O}(\delta\th^{n+1})$.
Now, since for $\th=0$  (\ref{her}) trivially  holds, we
conclude that (\ref{her}) holds for finite $\th$.

In particular if $\rho_\Psi$ is a unitary rep.:
$\rho_\Psi(\L)^\dagger=\rho_\Psi(\L)$,  then hermiticity
of $\rho_\Psi(A)$ implies hermiticity of  $\widehat{\rho_\Psi(A)}$
and of  $\widehat{\rho_\Psi(\L)}$.
\sk
\sk

In order to show reality of SW map, see (\ref{conj}), we complex conjugate 
(\ref{one}), (\ref{two}) and (\ref{three}) in the 
$\rho_\Psi$ representation and obtain
\eqa
&&\delta_\th\overline{\widehat{\rho_\Psi(A_\mu)}}
=-{1\over 4}\delta\th^{\rho\sigma}
\{{_{\,}}\overline{{\widehat{\rho_\Psi(A_\rho)}}}
{\;{}^{_{^{\textstyle \m*}}}
{_{\!{_{\!\!}}}\!\!\!\!\!,}}~~\;
\partial_\sigma\overline{\widehat{\rho_\Psi(A_\mu)}}+
\overline{\widehat{\rho_\Psi(F_{\sigma\mu})}}{_{\,}}\}
\label{61}\\
&&\delta_\th{\overline{\widehat{\rho_\Psi(\L)}}}
=-{1\over 4}\delta\th^{\rho\sigma}
\{{_{\,}}\partial_\rho{\overline{\widehat{\rho_\Psi(\L)}}}
{\;{}^{_{^{\textstyle \m*}}}
{_{\!{_{\!\!}}}\!\!\!\!\!,}}~~\;
\overline{\widehat{\rho_\Psi(A_\sigma)}}{_{\,}}\}
\label{62}\\
&&\delta_\th\overline{\Psih} =
-\frac{1}{2} \delta\theta^{\mu\nu}\overline{\widehat{\rho_\Psi(A_\mu)}}
{_{\,}}\m*
\pp\nu\overline{\Psih}\,
-\frac{i}{8}\delta\theta^{\mu\nu}[{_{_{\,}}}\overline{\widehat{\rho_\Psi(A_\mu)}}
{\;{}^{_{^{\textstyle \m*}}}
{_{\!{_{\!\!}}}\!\!\!\!\!,}}~~\;
\overline{\widehat{\rho_\Psi(A_\nu)}}{_{_{\,}}}]{_{\,}}
\m*
\overline{\Psih}
\label{63}
\ena
Comparison of (\ref{61}), (\ref{62}) and (\ref{63}) 
with SW differential equation for the  $\rho_\P*$ representation
\eqa
&&\delta_\th{\widehat{\rho_\P*(A_\mu)}}
=-{1\over 4}(-\delta\th)^{\rho\sigma}
\{{_{\,}}{{\widehat{\rho_\P*(A_\rho)}}}
{\;{}^{_{^{\textstyle \m*}}}
{_{\!{_{\!\!}}}\!\!\!\!\!,}}~~\;
\partial_\sigma{\widehat{\rho_\P*(A_\mu)}}+
{\widehat{\rho_\P*(F_{\sigma\mu})}}{_{\,}}\}
\label{71}\\
&&\delta_\th{{\widehat{\rho_\P*(\L)}}}
=-{1\over 4}(-\delta\th)^{\rho\sigma}
\{{_{\,}}\partial_\rho{{\widehat{\rho_\P*(\L)}}}
{\;{}^{_{^{\textstyle \m*}}}
{_{\!{_{\!\!}}}\!\!\!\!\!,}}~~\;
{\widehat{\rho_\P*(A_\sigma)}}{_{\,}}\}
\label{72}\\
&&\delta_\th\widehat{\P*} =
-\frac{1}{2} (-\delta\theta)^{\mu\nu}{\widehat{\rho_\P*(A_\mu)}}
{_{\,}}\m*{}_{\!}
\pp\nu\widehat{\P*}
\,+\frac{i}{8}(-\delta\theta)^{\mu\nu}{_{_{\,}}}[{_{_{\,}}}{\widehat{\rho_\P*(A_\mu)}}
{\;{}^{_{^{\textstyle \m*}}}
{_{\!{_{\!\!}}}\!\!\!\!\!,}}~~\;
{\widehat{\rho_\P*(A_\nu)}}{_{_{\,}}}]
{_{\,}}\m*
\widehat{\P*}~~~~~~~~~~{}
\label{73}
\ena
shows that if (\ref{conj}) holds at order ${\mathcal O}(\delta\th^n)$
then it is also true for 
$\widehat{\Psi}^{\!{'}}=\Psih+\delta_\th\Psih$, for
$\widehat{\rho_\Psi(A)}^{\!{_{{{\mbox{\normalsize{$'$}}}}}}}
=\widehat{\rho_\Psi(A)}+\delta_\th\widehat{\rho_\Psi(A)}$ and for
$\widehat{\rho_\Psi(\L)}^{\,{_{{{\!\!\mbox{\normalsize{$'$}}}}}}}=\widehat{\rho_\Psi(\L)}+\delta_\th\widehat{\rho_\Psi(\L)}\,$, 
i.e. (\ref{conj}) holds also  at order ${\mathcal O}(\delta\th^{n+1})$.
Now, since for $\th=0$  (\ref{conj})  holds, we
conclude that (\ref{conj}) holds for finite $\th$.

\sk
We end this section  observing that if  $\rho_\Psi$ is a unitary
representation, then the SW differential equation for  $\widehat{\P*}$ 
reads
\eq
\delta_\th\widehat{\P*} =
\frac{1}{2} \delta\theta^{\mu\nu}\pp\mu\widehat{\P*}
\*\widehat{\rho_\Psi(A_\nu)}+\frac{i}{8}\delta\theta^{\mu\nu}
\widehat{\P*}\*\s[\widehat{\rho_\Psi(A_\mu)} , \widehat{\rho_\Psi(A_\nu)}]~.
\label{80}
\en
Since the components of ${\Psih}^{\,\dagger}$ and 
${\widehat{\Psi}}^{{^{\,\mbox{\scriptsize{$*$}}}}}{\!}$  are the same,
comparison of (\ref{80}) with the hermitian conjugate of (\ref{three})
shows again that SW the map commutes with complex conjugation.

\sk
\sk
\noi{\bf 3.1$~~$
Gauge kinetic terms \mbox{\boldmath $\int T^{\!}r(\Fh\Fh)$} that are
invariant  under \mbox{\boldmath $\th\rightarrow - \th$}}
\sk
\noi
It is not difficult to derive from (\ref{conj}) the property
\eq
\overline{\widehat{\rho_{\Psi^{}}(F)}}=-\widehat{\rho_{\Psi^{*}}(F)}
\en 
where (as in (\ref{27}) and (\ref{29})) we recall that on the l.h.s., the SW
map with $+\th$ is used, while the SW map with $-\th$ enters the r.h.s..
Reality of the gauge kinetic term then implies
\eq
\int\!d^4x\; \Tr(\widehat{\rho_{\Psi^{}}(F)}\,\widehat{\rho_{\Psi^{}}(F)})
=
\int\!d^4x\; \Tr(\overline{\widehat{\rho_{\Psi^{}}(F)}}\;
\overline{\widehat{\rho_{\Psi^{}}(F)}})
=
\int\!d^4x\; \Tr(\widehat{\rho_{\Psi^{*}}(F)}\,\widehat{\rho_{\Psi^{*}}(F)})
\label{realr}\en 
where in the last expression the SW map with $-\th$ is used.
We thus see that the gauge kinetic term $\int T^{\!}r(\Fh\Fh)$ can be associated 
with $+\th$ or $-\th$ depending on the representation used.
In particular for representations that are real (i.e. $\rho=\rho^*$ up
to a similarity transformation) equality (\ref{realr}) implies
that the gauge kinetic term is even in $\th$. 
An important example of real representations is given by the adjoint representation.

\section{$C$, $P$ and $T$}
We consider the noncommutativity parameter $\th$ as a two-tensor
that transforms covariantly under Lorentz rotations and more 
generally under $C,P$ and $T$. These transformation properties of $\th$
are compatible with the relation \cite{chu,SW} between $\th$,
the closed string metric $g$ and the NS $B$-field 
(that transforms as a field strength)
$
\th=({1\over g+B})_{{_{\!A}}}
$
where $(~~)_{{_{\!A}}}$ denotes the antisymmetric part of the matrix.
Under time inversion and parity we explicitly have
$~\Lambda ~{\stackrel{{}_{_{\scriptstyle T}}}{-\!\!\!
\longrightarrow}}~
{\Lambda^T}={\Lambda}\,\,$,
$\,\,\Lambda ~{\stackrel{{}_{_{\scriptstyle P}}}{-\!\!\!
\longrightarrow}}~
{\Lambda^P}={\Lambda}\,\,$ and\footnote{We use two component spinor
notation and the  Weil representation
$\gamma^0=~\!\!
\left( {}^{\mbox{\sma {$0$}} 
\mbox{\sma {$^{}\;\1_{}$}}}_{\mbox{\sma {$\1$}} 
\mbox{\sma{$\;0_{}$}}} \right)$,
$\,\gamma^i=~\!\!
\left( {}^{\mbox{\sma {$~~0$}} 
\mbox{\sma {$~~~\!\sigma^i$}}}_{\mbox{\sma {$-\sigma^i$}} 
\mbox{\sma{$\,\,0$}}} \right)$, 
$\,\gamma^5=~\!\!
\left( {}^{\mbox{\sma {$-\1$}} 
\mbox{\sma {$\,\,0$}}}_{\mbox{\sma {$~\,\,\,0$}} 
\mbox{\sma{$~\,\1$}}} \right)$. We also write 
$\,\gamma^\mu=~\!\!
\left( {}^{\mbox{\sma {$\;0$}} 
\mbox{\sma {$~~~\!\sigma^\mu$}}}_{\mbox{\sma 
{$\overline{\sigma}^\mu$}} 
\mbox{\sma{$~\;0$}}} \right)$. 
}
\eqa
&&
{}~~~\th^{\mu\nu}~{\stackrel{T}{-\!\!\!-\!\!\!-\!\!\!
\longrightarrow}}~
{\th^T}^{\,{\mu\nu}}=
\left
\{{}^{{}^{{}{\textstyle{~\;\th^{0j}}}}}_{{}_{{}{\textstyle{-\th^{ij}}}}}
\right.~~~~~,~~~~~~\!\!
A_{\mu}~{\stackrel{T}{-\!\!\!-\!\!\!-\!\!\!
\longrightarrow}}~
{A^T}_{{\!\!\mu}}=
\left
\{{}^{{}^{{}{\textstyle{~\;A_{0}}}}}_{{}_{{}{\textstyle{-A_{i}}}}}
\right.~\label{AT}\\[.6em]
&&
{}^{{}^{{}{\textstyle{~~\;
\Psi_{\!L}~\;{\stackrel{T}{-\!\!\!-\!\!\!-\!\!\!
\longrightarrow\;}}~
{\Psi_{\!L}^{\,T}}={-i\sigma_1\sigma_3\Psi_{\!L}}
}}}}_{{}_{{}{\textstyle{~~~
\Psi_{\!R}~\;{\stackrel{T}{-\!\!\!-\!\!\!-\!\!\!
\longrightarrow\;}}~
{\Psi_{\!R}^{\,T}}={-i\sigma_1\sigma_3\Psi_{\!R}}
}}}}
~~~\!\!\!,\,~~~~~~
\partial_{\mu}~{\stackrel{T}{-\!\!\!-\!\!\!-\!\!\!
\longrightarrow}}~
{\partial^T}_{{\!\!\mu}}=
\left
\{{}^{{}^{{}{\textstyle{-\partial_{
0}}}}}_{{}_{{}{\textstyle{~\;\partial_{i}}}}}\right.\label{psiT}
\ena
{\vskip .2em}
\eqa
&&
{}~~~\th^{\mu\nu}~{\stackrel{P}{-\!\!\!-\!\!\!-\!\!\!
\longrightarrow}}~
{\th^P}^{\,{\mu\nu}}=
\left
\{{}^{{}^{{}{\textstyle{-\th^{0j}}}}}_{{}_{{}{\textstyle{~\;\th^{ij}}}}}
\right.~~\;~~,~~~~~~
A_{\mu}~{\stackrel{P}{-\!\!\!-\!\!\!-\!\!\!
\longrightarrow}}~
{A^P}_{{\!\!\mu}}=
\left
\{{}^{{}^{{}{\textstyle{~\;A_{0}}}}}_{{}_{{}{\textstyle{-A_{i}}}}}
\right.\label{Aparity}\\[.4em]
&&
{}^{{}^{{}{\textstyle{~~\;
\Psi_{\!L}~\;{\stackrel{P}{-\!\!\!-\!\!\!-\!\!\!
\longrightarrow\;}}~
{\Psi_{\!L}^{\;P}}={\Psi_{\!R}}
}}}}_{{}_{{}{\textstyle{~~~
\Psi_{\!R}~\;{\stackrel{P}{-\!\!\!-\!\!\!-\!\!\!
\longrightarrow\;}}~
{\Psi_{\!R}^{\;P}}={\Psi_{\!L}}
}}}}~~~~~~~~~~\,,\!\!\!\!~~~~~~~~~
\partial_{\mu}~{\stackrel{P}{-\!\!\!-\!\!\!-\!\!\!
\longrightarrow}}~
{\partial^P}_{{\!\!\mu}}=
\left
\{{}^{{}^{{}{\textstyle{~\,\partial_{
0}}}}}_{{}_{{}{\textstyle{-\partial_{i}}}}}\right.\label{psiparity}
\ena
Under charge conjugation we have\footnote{Charge conjugation does not
act on the Lorentz structure, therefore it maps  left (right) 
handed fermions to  left (right) handed fermions.}
\eq
\th^{\mu\nu}~{\stackrel{C}{-\!\!\!-\!\!\!-\!\!\!
\longrightarrow}}~
{\th^C}^{\,{\mu\nu}}=-\th^{\mu\nu}
\label{th*}
\en
{\vskip -2em}
\eqa
&&
A_{\mu}=A_{\mu}^aT^a~{\stackrel{C}{-\!\!\!-\!\!\!-\!\!\!
\longrightarrow}}~
{A^{C}}_{{\!\!\!\mu}}=-\overline{A_\mu}
~~~~~~~~~~~,~~~
\L=\L^aT^a~{\stackrel{C}{-\!\!\!-\!\!\!-\!\!\!
\longrightarrow}}~
{\L^{C}}_{{\!\!\!\mu}}=-\overline{\L}
\label{ACC}\\[.2em]
&&
\Psi_{\!L}~\;{\stackrel{C}{-\!\!\!-\!\!\!-\!\!\!
\longrightarrow\;}}~
{\Psi_{\!L}}^{\,C}={-i\sigma_2{_{\,}}\Psi_{\!R}^{\,*}}={\Psi_{~L}^{_{\!}C}}
~~~~~\,,~~~
\Psi_{\!R}~\;{\stackrel{C}{-\!\!\!-\!\!\!-\!\!\!
\longrightarrow\;}}~
{\Psi_{\!R}^{\;C}}={i\sigma_2{_{\,}}}\Psi_{\!L}^{\,*}
={\Psi_{~R}^{_{\!}C}}
~.~~~~~~~~~~\label{PsiCC}
\ena
where $\overline{A_\mu}=\overline{A_\mu^a}\,\overline{T^a}
=A_\mu^a\overline{T^a}\,$ and
 $\;\overline{\L}=\overline{\L^a}\,\overline{T^a}=\L^a\overline{T^a}$
are shorthand notations for the complex conjugate representation.
More precisely 
\eq
\rho_{\Psi_{\!L}}(T^a) ~\;{\stackrel{C}{-\!\!\!-\!\!\!-\!\!\!
\longrightarrow\;}}~
(\rho_{\Psi_{\!L}}(T^a))^C
=-\overline{\rho_{\Psi_{\!L}}(T^a)}=\rho_{\PL*}(T^a)~,
\label{LambdaCC}
\en
where we used 
(\ref{commU}). Similarly 
$\rho_{\Psi_{\!R}}(T^a) \;{\stackrel{C}{\longrightarrow\;}}
-\overline{\rho_{\Psi_{\!R}}(T^a)}=\rho_{\Psi_{\!R}^{*}}(T^a)\,.$
\sk
Let us consider the SW map 
$$
{\widehat{\Psi_{\!L}}}
=\SW[\Psi_{\!L},
\rho_{\Psi_{\!L}} (A),
\th,\partial,i]
$$
where we have written explicitly also the dependence on the partial
derivatives; the imaginary unit $i$ in the last slot marks that the coefficients in the
SW map are in general complex coefficients.
Since $T$ is antilinear and multiplicative we have
\eq
{\widehat{\Psi_{\!L}}^{\:T}}
=\SW[\Psi_{\!L}^{{^{^{\,}}}T},
\rho_{\Psi_{\!L}^{{^{^{\,}}}T}}(A^T),
\th^T,\partial^T,-i]
\label{Top}
\en
where now $-i$ means that we are considering the complex conjugates
of the coefficients in the SW map.
Similarly 
\eq
\Ah^{\textstyle{^{\;T}}}=\Ah[A^T,\th^T,\partial^T,-i]~~,~~~
\lh^{\textstyle{^{\;T}}}=\lh[\Lambda^T,A^T,\th^T,\partial^T,-i]~~~~
\en
more precisely  we should write $
{\widehat{\rho_{\Psi_{\!L}}(A)}}^T=
\Ah[\rho_{\Psi_{\!L}^{}}(A^T),\th^T,\partial^T,-i]$
and similarly for $\lh$.

We now show that the 
$T$ operation  on hatted variables has the same expression as on
unhatted variables:
\eqa
{\widehat{\Psi_{\!L}}^{\:T}}={-i\sigma_1\sigma_3\widehat{\Psi_{\!L}}}
~~&,~~
\Ah_{\,\mu}^{\textstyle{^{\;T}}}= 
\left
\{{}^{{}^{{}{\textstyle{~\;\Ah_{0}}}}}_{{}_{{}{\textstyle{-\Ah_{i}}}}}
\right.~~~,&~~\lh^{\textstyle{^{\;T}}}= \lh   ~.     
\label{LT}
\ena
We first notice that (\ref{conj}) implies (use (\ref{commU}) and
replace $\Psi^*$ with $\Psi_{\!L}$) 
\eq
\SW[\Psi_{\!L},
\rho_{\Psi_{\!L}}(A),
\th,\partial,-i]=
\SW[\Psi_{\!L},
-\rho_{\Psi_{\!L}}(A),
-\th,\partial,i]
\label{70}
\en
\eq
\Ah[
\rho_{\Psi_{\!L}}(A),
\th,\partial,-i]=
-\Ah[-\rho_{\Psi_{\!L}}(A),
-\th,\partial,i]
\label{771}
\en
and similarly for $\lh$.
We then have
\eq
{\widehat{\Psi_{\!L}}^{\:T}}=
\SW[-i\sigma_1\sigma_3{\Psi_{\!L}},
-\rho_{\Psi_{\!L}}(A_0),\rho_{\Psi_{\!L}}(A_i),
-\th^{0i},\th^{ij},-\partial_0,\partial_i,i]
=-i\sigma_1\sigma_3{\widehat{\Psi_{\!L}}}
\label{plt}
\en
where in the last passage we factoried 
$-i\sigma_1\sigma_3$ and noticed that the $-$ signs
appear together with the index $0$ and that since the SW map
preserves the space-time index structure, the $-$ signs
appear always in pairs. One proceeds similarly with 
$\Ah^{\textstyle{^{\;T}}}$ and $\lh^{\textstyle{^{\;T}}}$.
\sk
In order to discuss parity and charge conjugation on 
noncommutative spinors we first consider 
noncommutative QED with just a $4$-component Dirac spinor $\psi$, and
decompose it into its Weil spinors $\psi_L$ and $\psi_R$.
Their charge conjugate spinors are
$\psi^{\;C}_{\!L}=\psi^C_{~L}=-i\sigma_2\psi_R^{\,*}$
and  $\psi^{\;C}_{\!R}=\psi^C_{~R}=i\sigma_2\psi_L^{\,*}$. 
Once we define 
$$\widehat{\psi_L}=\SW [\psi_L,\rho_{\psi_L}(A),\th,\partial,i]~,$$ 
we then have the choice 
\eq
\widehat{\psi_R}=\SW [\psi_R,\rho_{\psi_R}(A),\mbox{\boldmath ${\pm}$}
\th,\partial,i]~.
\label{choice}
\en
In the literature the choice $+\th$ is usually
considered so that for the $4$-component Dirac spinor $\psi$ we can
write  $\widehat{\psi}=\SW [\psi,A,\th,\partial,i]$, 
$\delta\widehat{\psi}=i\widehat{\Lambda}\*\widehat{\psi}$.
With this choice, the gauge potential $A$ and the noncommutativity
parameter $\th$ appear with the same sign in 
$\widehat{\psi_L}$ and  $\widehat{\psi_R}$.
We here advocate the opposite choice ($-\th$) in (\ref{choice}).
Indeed we have that 
\eq
\widehat{\psi_R}=\SW [\psi_R,\rho_{\psi_R^{}}(A),-\th,\partial,i]~~~
\Longleftrightarrow~~~
\widehat{\psi_{\!L}^{\;C}}=\SW [{\psi_{\!L}^{\;C}},
\rho_{\psi_{\!L}^{\;C}}(A),+\th,\partial,i]
\label{equiv}
\en
so that with the $-\th$ choice in (\ref{choice}), 
both left handed fermions $\widehat{\psi_L}$,
$\widehat{{\psi^{\;C}_{\!L}}}$
are associated with $\th$ while the right handed ones $\widehat{\psi_R}$, 
$\widehat{{\psi^{\;C}_{\!R}}}$ are associated with $-\th$.
In GUT theories we have multiplets of definite chirality
(see e.g.(\ref{ml})) and therefore this is the natural choice to
consider in this setting. Property (\ref{equiv}) is easily proven (recall (\ref{conj})
with $-\th$ instead of $\th$)
\eqa
\widehat{\psi_{\!L}^{\;C}}&=&\SW [-i\sigma_2{\psi_{\!R}^{\;*}},
\rho_{\psi_{\!R}^{\;*}}(A),\th,\partial,i]
=-i\sigma_2\,\SW [{\psi_{\!R}^{\;*}},{\rho_{\psi_{\!R}^{*}}(A)},\th,\partial,i]\nonumber\\
&=&-i\sigma_2\,\overline{\SW [{\psi_{\!R}^{}},
\rho_{\psi_{\!R}^{}}(A),-\th,\partial,i]}
=-i\sigma_2\,{\widehat{\psi_R}}^{\,*}~.
\label{ultim}
\ena
In the following, in order to describe both $\pm\th$ choices, we write
\eq
{\widehat{\Psi_{\!R}}}
=\SW[\Psi_{\!R},
\rho_{\Psi_{\!R}} (A),
\sr(\th),\partial,i]
\en
where 
$~\sr (\th)=\pm\th~$
depending on (\ref{choice}).
The $T,P,C$ transformed spinors then read 
\eq
\widehat{\Psi_{\!R}}^{\,T}=
\SW[\Psi_R^{\,T},\rho_{\Psi_{_{\!}R}}(A^T),
\sr (\th^T),\partial^T,-i]
\label{PT}
\en
and 
\eqa
&\widehat{\Psi_{\!L}}^{\,P}=
 \SW[\Psi_{\!L}^{\,P},\rho_{\Psi_{_{\!}L}}^{}(A^P),\th^P,\partial^P,i]~,&~
{\widehat{\Psi_{\!R}}^{\:P}}=
\SW[\Psi_{\!R}^{{^{^{\,}}}P},\rho_{\Psi_{\!R}^{{^{{}}}}}(A^P),
\sr(\th^P),\partial^P,i]~~~
\label{Pop}\\
&\widehat{\Psi_{\!L}}^{\,C}=
\SW[\Psi_{\!L}^{\,C},{(\rho_{\Psi_{_{\!}L}^{}}(A))^C},
\th^C,\partial,i]~,&~
{\widehat{\Psi_{\!R}}^{\:C}}=
\SW[\Psi_{\!R}^{\,C},
{(\rho_{\Psi_{\!R}^{}}(A))^C},\sr(\th^C),\partial,i]~~~
\label{Cconj}
\ena
Consistently with (\ref{PT})-(\ref{Cconj}) we also have 
${\widehat{\rho_{\Psi_{\!R}}(A)}}^T\!=
\Ah[\rho_{\Psi_{\!R}^{}}(A^T),\sigma_{\Psi_{\!R}}(\th^T),\partial^T,-i
]$ and 
\eq
{\widehat{\rho_{\Psi_{\!L}}(A)}}^P\!=
\Ah[\rho_{\Psi_{\!L}^{}}(A^P), \th^P,\partial^P,i]~,~~
{\widehat{\rho_{\Psi_{\!L}}(A)}}^C\!=
\Ah[(\rho_{\Psi_{\!L}^{}}(A))^C,\th^C,\partial,i]
\label{mth}
\en
\eq
{\widehat{\rho_{\Psi_{\!R}}(A)}}^P=
\Ah[\rho_{\Psi_{\!R}^{}}(A^P),\sr(\th^P),\partial^P,i]~~,~~~~
{\widehat{\rho_{\Psi_{\!R}}(A)}}^C=
\Ah[(\rho_{\Psi_{\!R}^{}}(A))^C,\sr(\th^C),\partial,i]
\label{LR}
\en
and similarly for $\lh$.

\sk
If we replace $\th$ and $\Psi_{\!L}$ with 
$\sigma_{\Psi_{\!R}^{}}(\th)$ and $\Psi_{\!R}^{}$ in 
(\ref{plt}) we immediately have that the $T$ transformation on right 
handed hatted variables has the same expressions as on unhatted variables
\eqa
{\widehat{\Psi_{\!R}}^{\:T}}={-i\sigma_1\sigma_3 \widehat{\Psi_{\!R}}}~
~~&,~~
\widehat{\rho_{\Psi_{\!R}^{}}(A_{\,\mu})}^{\textstyle{^{\;T}}}= 
\left
\{{}^{{}^{{}{\textstyle
{~\;\widehat{\rho_{\Psi_{\!R}^{}}(A_{\,0})}}}}}_{{}_{{}
{\textstyle{-\widehat{\rho_{\Psi_{\!R}^{}}(A_{\,i})}}}}}
\right.~~~,&~~\widehat{\rho_{\Psi_{\!R}^{}}(\Lambda)}^{\textstyle{^{\;T}}}
= \widehat{\rho_{\Psi_{\!R}^{}}(\Lambda)}   ~.     
\label{RT}
\ena
\sk
We now show that for the $\mbox{\boldmath ${+\th}$}$ choice 
of equation (\ref{choice}) parity and charge conjugation on  hatted variables have the same 
expressions as on unhatted variables:
\eqa
{\widehat{\Psi_{\!L}}^{\:P}}={\widehat{\Psi_{\!R}}}
      ~~~~~~~~~~,&~~~~
{\widehat{\Psi_{\!R}}^{\:P}}={\widehat{\Psi_{\!L}}}~~&,~~  
\label{PsiC&P&T}\\
\Ah_{\,\mu}^{\textstyle{^{\;P}}}=
\left
\{{}^{{}^{{}{\textstyle{~\;\Ah_{0}}}}}_{{}_{{}{\textstyle{-\Ah_{i}}}}}
\right.~~~~~~,&~~~~
{\lh}^{\textstyle{^{\;P}}}= \lh                   \; ~~~&,~~
\label{LPar}\\
{\widehat{\Psi_{\!L}}^{\:C}}={-i\sigma_2\widehat{\Psi_{\!R}}^*}
      ~~\,,&~~~~~~~~~~
{\widehat{\Psi_{\!R}}^{\:C}}={i\sigma_2{\widehat{\Psi_{\!L}}}^*} ~\;~&,~~
\label{PsiC&P}\\
\Ah^{\textstyle{^{\;C}}}= -\overline{\widehat{A}}
       ~~~~~~~~~~,&~~~~
\lh^{\textstyle{^{\;C}}}= -\overline{\widehat{\L}} ~\;&.~~
\label{LCon}
\ena
Under parity  we have
\eq
\widehat{\Psi_{\!L}}^{\,P}\!=
 \SW[\Psi_{\!R}^{},\rho_{\Psi_{{\!}L}^{}}(A^P),\th^P,\partial^P,i]
=\SW[\Psi_{\!R},\rho_{\Psi_{\!R}^{}}(A),\th,\partial,i]
=\widehat{\Psi_{\!R}}
\label{SWP}\;
\en
where in the second equality we removed the apex ${}^P$ because the SW map
preserves the space-time index structure, we also used 
$\rho_{\Psi_{\!L}^{}} =\rho_{\Psi_{\!R}^{}}$. If this condition is not
met then parity (for $\th=0$ and henceforth for $\th\not=0$) 
is surely broken, and the expression 
$\widehat{\Psi_{\!L}}^{\,P}\!$ is usually 
meaningless.  As in (\ref{SWP}) we also have
$
\widehat{\Psi_{\!R}}^{\,P}=\widehat{\Psi_{\!L}}\,.$

The proof of (\ref{LPar}) again relies on the space-time index
structure of $\Ah$ and $\lh$.
\sk
We proceed similarly in the case of $C$. For example 
reality of the SW map (\ref{conj}) leads to
\eq
\widehat{\Psi_{\!L}}^{\,C}
=\SW[\Psi_{\!L}^{\,C},-\overline{\rho_{\Psi_{_{\!}L}^{}}(A)},\th^C,\partial,i]
=-i\sigma_2\,\SW[\Psi_{\!R}^{\,*},\rho_{\Psi_{\!R}^{*}}(A),-\th,\partial,i]
\label{last}
=-i\sigma_2{_{\,}}\widehat{\Psi_{\!R}}^{\,*}\nonumber~
\en
where we again used $\rho_{\Psi_{\!L}^{}} =\rho_{\Psi_{\!R}^{}}$,
a necessary condition for charge conjugation symmetry.
As in (\ref{last}) we also have  $\widehat{\Psi_{\!R}}^{\,C}=
i\sigma_2\widehat{\Psi_{\!R}}^*$. The proof of (\ref{LCon})
easily follows from (\ref{conj}).
\sk
In the  $\mbox{\boldmath ${-\th}$}$ case of (\ref{choice}) the equalities
(\ref{PsiC&P&T})-(\ref{LCon}) do not hold because 
$\th$ appears with the wrong sign. We can cure this by defining 
$\th^P$ and $\th^C$ with an extra $-$ sign 
(i.e. $\th^{P\;{ij}}=-\th^{ij},\; \th^{P\;{0i}}=\th^{0i}$ and $\th^C=\th$)
then (\ref{PsiC&P&T})-(\ref{LCon}) hold.
\sk
Finally notice that independently from the $\pm\th$ choice (and from
the $C_{\!}P$ symmetry of commutative actions) we can always consider
the $C_{\!}P$ transformed SW map, and we have
\eq
{\widehat{\Psi_{\!L}}^{\:C\!P}}=i\sigma_2{\widehat{\Psi_{\!L}}}^{*}
      ~~~,~~~~
{\widehat{\Psi_{\!R}}^{\:C\!P}}=-i\sigma_2{\widehat{\Psi_{\!R}}}^{*} 
~~~,~~~~\Ah_{\,\mu}^{\textstyle{^{\;C\!P}}}=
\left
\{{}^{{}^{{}{\textstyle{-\overline{\Ah_{0}}}}}}_{{}_{{}
{\textstyle{~\;\overline{\Ah_{i}}}}}}
\right.~~~,~~~~
{\lh}^{\textstyle{^{\;C\!P}}}= -\overline{\lh}    ~~.~~~
\en
We have concentrated on the case of constant theta in 
this section but 
the results should still be valid in the general $\theta(x)$ case.
To show this one needs to consider the methods of \cite{Jurco}
in place of the SW differential equation (which is limited
to the Moyal-Weyl star product).

\subsection{Noncommutative QED$_+$ and QED$_-$}
QED$_\pm$ are the two different QED theories obtained with the two
different $\pm\th$ choices (\ref{choice}). It is easy to compare the
two constructions. We have (up to gauge kinetic terms)
\eq
S_\QEDp = \int \overline{\widehat{\psi_{}}}\* i
\widehat{\fmslash D} \widehat \psi_{}=\int {\widehat{\psi_{\!L}}}^\dagger\* i
\widehat{\fmslash D}_{\,} \widehat \psi_{\!L}
\,+\,{\widehat{\psi_{\!R}}}^\dagger\* i
\widehat{\fmslash D}_{\,} \widehat \psi_{\!R}~.
\en
On the other hand, the GUT inspired $\QEDm$ is obtained 
considering the left handed spinor
\(\chi_L^{}=\left({^{^{^{\!}}}}{_{_{_{\!\!}}}}\right.
{}^{^{\mbox{\sma{$\psi_{\!L}^{} $}}}}_{_{\mbox{\sma{$\psi^C_{\!L} $}}}}
\left.{^{^{^{\!}}}}{_{_{_{\!\!\!}}}}\right)\)
so that $\widehat{\psi^C_{\!L}}
=\SW[\psi^C_{\!L},\rho_{\psi^C_{\!L}}(A),\th,\partial,i]$. 
We have
\eq
S_\QEDm =\int {\widehat{\psi_{\!L}}}^\dagger\* i
\widehat{\fmslash D}_{\,} \widehat \psi_{\!L}
\,+\,{\widehat{\psi^{\;C}_{\!L}}}^\dagger\* i
\widehat{\fmslash D}_{\,} \widehat{\psi^{\;C}_{\!L}}\:.
\en
Now, from (\ref{ultim}) and $\sigma$ matrices algebra we have
$\,\int {\widehat{\psi^{\;C}_{\!L}}}^\dagger\* i
\widehat{\fmslash D}_{\,} \widehat{\psi^{\;C}_{\!L}}=
\int {\widehat{\psi_{\!R}}}^{\!{{
\mbox{${_{^{{{o{^{_{^{_{\!\!}}}}}p}}}}}$}}}}
{}{^{\mbox{${}^\dagger$}}}\m* i
\widehat{\fmslash D}^{\!{{
\mbox{${_{^{{{^{\:{_{\!\!\!}}}}{o{^{_{^{_{\!\!}}}}}p}}}}}$}}}}
_{\,} \widehat{\psi_{\!R}}^{\!{{
\mbox{${_{^{{{o{^{_{^{_{\!\!}}}}}p}}}}}$}}}}
$,
where we have emphasized  that we are using the $-\th$ convention
in the SW map by writing 
${~}\widehat{^{}}^{\!{{\mbox{${_{^{{\,\;{o{^{_{^{_{\!\!}}}}}p}}}}}$}}}}$ 
instead of ${~}\widehat{^{}}{~\,}$. 
We conclude that in order to obtain QED$_-$ from QED$_+$ we just need to 
change $\th$ into $-\th$ in the right handed fermion sector of
QED$_+$.

\section{$C,P,T$ properties of NCYM actions.}

\noi In this section we derive the transformations properties
of NCYM actions. We assume that $\th$ transforms as in 
(\ref{AT}), (\ref{Aparity}) and (\ref{ACC}). Then  in  the $+\th$
choice (\ref{choice}) we have that NCYM actions are 
invariant under $C,P$ and $T$ iff in the commutative limit they are 
invariant. On the other hand, in the $-\th$ choice
NCYM actions are invariant under $C_{\!}P$ and $T$ iff in the commutative 
limit they are invariant.
For the fermion kinetic term these statements are
a straighforward consequence of $\int {\widehat{\Psi_{\!L}}}^\dagger\*
{\fmslash \partial}{\widehat{\Psi_{\!L}}}
=\int {\widehat{\Psi_{\!L}}}^\dagger
{\fmslash \partial}{\widehat{\Psi_{\!L}}}$ (and similarly for 
${\widehat{\Psi_{\!R}}}$).
Since $\Fh$ transforms like $F$ under $C_{\!}P$ and $T$, and in the
$+\th$ case also under $C$ and $P$ separately,\footnote{
$\Fh$ transforms like $F$ under $P$ and $T$ because $P$ and $T$ leave invariant
the $\*$-product (indeed $(i\th^{\mu\nu}{\stackrel{\leftarrow}{\partial}_{\!\mu}}
{\stackrel{\rightarrow}{\partial}_{\!\nu}})^T=-i\,
(\th^{\mu\nu}{\stackrel{\leftarrow}{\partial}_{\!\mu}}
{\stackrel{\rightarrow}{\partial}_{\!\nu}})^T=
i\th^{\mu\nu}{\stackrel{\leftarrow}{\partial}_{\!\mu}}
{\stackrel{\rightarrow}{\partial}_{\!\nu}}$).
Under charge conjugation  we also have $\Fh^{^{\scriptstyle C}}=-\overline{\Fh}\,,$ indeed
$i\s[\Ah^C_{},{}^{^{\!\!\!C}}\Ah^C_{}]=-_{\,}\overline{_{\,}i\s[\Ah_{},\Ah_{}]}$
because the action of $C$ on $\*$ equals complex
conjugation.}
the $C,P$,$T$ properties of the gauge kinetic term 
$\int T\!r(\Fh\*\Fh)=\int T\!r(\Fh\Fh)$ easily follow.
Inspection of the fermion gauge bosons interaction term
leads also to the same conclusion. 
For sake of clarity we treat separately the $+\th$ choice and 
the $-\th$ choice (cf. (\ref{choice})).
\sk
\noi {$\mbox{\boldmath ${+ \theta}$}$ {\bf{case}}} $~$ For a $4$-component Dirac spinor 
the interaction term is  $\overline{\psih}\*\gamma^{\mu}\Ah_\mu\*\psih$.
Invariance under  $P$ and $T$ transformations is straighforward
since $P$ and $T$ do not change the $\*$ product.
Invariance under charge conjugation follows from 
\eqa
(\overline{\psih}\*\gamma^{\mu}\Ah_\mu\*\psih)^{^{{\scriptstyle C}}}
&\equiv&
\overline{\psih^{^{{\:\scriptstyle C}}}}\m*\gamma^{\mu}
\Ah_\mu^{^{{\:\scriptstyle C}}}
{_{\,}}\m*
\psih^{^{{\:\scriptstyle C}}}
= {i(\gamma^0\gamma^2\psih)}^t\m*\gamma^{\mu}
(-\Ah_\mu)^t
\m*
i(\overline{\psih}\gamma^0\gamma^2)^t\nonumber\\
&=&\left({}^{^{}}_{_{}}\right.\!
\overline{\psih}\*\gamma^{\mu}\Ah_\mu\*\psih \!\left.{}^{^{}}_{_{}}
\right)\!^t
=\overline{\psih}\*\gamma^{\mu}\Ah_\mu\*\psih
\nonumber
\ena
where we used hermiticity of $\Ah$, the standard gamma matrix algebra 
and, as usual, that spinors anticommute. 

\sk
\noi {$\mbox{\boldmath ${- \theta}$}$ {\bf{case}}} $~$ 
In two components notation we have the interaction terms
\eq
{\widehat{\Psi_{\!L}}}^\dagger\*{\fmslash \Ah}\*
{\widehat{\Psi_{\!L}}}\,+\,
{\widehat{\Psi_{\!R}}}^\dagger\m*{\fmslash \Ah}{_{\,}}{_{\,}}\m*
{\widehat{\Psi_{\!R}}}\nonumber
\label{lrlr}
\en
Notice that $\widehat{\Psi_{{\!}R}}$ (consistently with (\ref{choice})) 
commands the opposite star product $\m*$.
Invariance under time reversal is straighforward since $T$ 
leaves invariant the $\*$ and $\m*$ products.
Under $C_{\!}P$ we have 
\eqa
\left({}^{^{}}_{{_{}}}\!\right.\widehat{\Psi_{\!L}}^\dagger\*{\fmslash\Ah}
\*\widehat{\Psi_{\!L}}\!\left.{}^{^{}}_{\!{_{}}}\right)^{C\!P}\!
&=&
\widehat{\Psi_{\!L}}^\dagger
{}^{^{^{\scriptstyle C\!P}}}
\!\!\*^{C\!P}\,\overline{\sigma}{_{\,}}
\Ah^{^{\,\scriptstyle C\!P}}\!\*^{C\!P}\widehat{\Psi_{\!L}}^{{
\scriptstyle C\!P}}
\!=\left({}^{^{}}_{\!{_{}}}\!\right.
i\sigma_2{\widehat{\Psi_{\!L}}}^{*}\left.{}^{^{}}_{\!{_{}}}\!\right)^\dagger
{_{\!}}\m*_{\,}
\overline{\sigma}{_{\,}}
\left({}^{^{}}_{\!{_{}}}\!\right.
-\overline{\Ah}
\left.{}^{^{}}_{\!{_{}}}\right)^P
\m*
i\sigma_2{\widehat{\Psi_{\!L}}}^{*}
\nonumber\\
&=&{\widehat{\Psi_{\!L}}}^ti\sigma_2
\m*
\sigma\,\Ah^{^{\,\scriptstyle t}}
\m*
i\sigma_2{\widehat{\Psi_{\!L}}}^*
=-{\widehat{\Psi_{\!L}}}^t
\m*
\overline{\sigma}^t\Ah^{^{\,\scriptstyle t}}
\m*
{\widehat{\Psi_{\!L}}}^*\nonumber\\
&=&{\widehat{\Psi_{\!L}}}^\dagger
\*
{\fmslash\Ah}
\* 
{\widehat{\Psi_{\!L}}}
\ena
Similarly for ${\widehat{\Psi_{\!R}}}$.
\sk\sk
We have studied the $C,P$ and $T$ symmetry properties of NCYM 
actions where the $\th$ transformations under $C,P$ and $T$ 
are given in (\ref{Aparity}),(\ref{AT}) and (\ref{ACC}).
Viceversa, {\sl if  we keep $\th$ fixed} under $C,P$ and $T$ 
transformations, we in general have that NCYM theories 
{\sl break}  $C,P$ and $T$ symmetries. 
Notice however that in the $-\th$ case, if we keep  $\th$ fixed
under $C$, then the SW map is well behaved under $C$, and $C$ is a symmetry 
of a NCYM action if it is a symmetry of the corresponding commutative one
and the gauge kinetic term $\int T^{\!}r(\Fh\Fh)$ is even in $\th$.
For example one can check that when
$\rho_{\Psi_{\!L}}=\rho_{\Psi_{\!R}}$,
the sum in (\ref{lrlr}) is invariant
under $C$.

\sk
Finally,
from (\ref{Aparity}),(\ref{AT}) (\ref{ACC}) (and (\ref{psiT}), 
(\ref{psiparity}), (\ref{PsiCC})) it follows that
under the combined $CPT$ transformation, $\th$ does not change, 
and therefore $CPT$ (with fixed $\theta$) 
is always a  symmetry of NCYM actions. 
In models where $\theta$ changes sign under $CPT$,
e.g., for nonconstant $\theta^{\mu\nu} = C^{\mu\nu}_\rho x^\rho$
with fixed background $C^{\mu\nu}_\rho$,  we do expect
spontaneous breaking of $CPT$.

\sk
\sk

\noi \textbf{\itshape cpt breaking}

\noi Here we do not consider the $CPT$ operator, but the
$cpt$ one. $CPT$ and $cpt$ differ only in their
action on $\th$, we have $\th^{cpt}=-\th$ (while $\th^{CPT}=\th$).
In particular, in the commutative case $\th=0$, we have $CPT=cpt$.
The transformation $\th^{cpt}=-\th$ can be justified by a 
quantum mechanics analogy. In QM the antiunitary $cpt$ 
operator acts on
the ${\bx}^i$ and ${\bp}^i$ operators via conjugation so that
$cpt (\bx {\scriptstyle\circ} \bp) cpt^{-1}=
cpt_{_{_{}}} \bx_{_{_{}}} cpt^{-1}\,{\scriptstyle\circ}\, cpt_{_{_{}}} 
\bp_{_{_{}}} cpt^{-1}=-\bx_{_{}} {\scriptstyle\circ} \bp$, (and the 
$[\bx^i\,,{\!\!\!^{_{_{\scriptstyle{\circ}}}}}\bp^j] =i\hbar\delta^{ij}$ 
relations are invariant under $cpt$). It is then natural to define
$cpt (\bx^{\mu} {\scriptstyle\circ}_{\,} \bx^{\nu}) cpt^{-1}=
cpt_{_{_{}}} \bx^\mu cpt^{-1}\,{\scriptstyle\circ}\, cpt_{_{_{}}} 
\bx^\nu cpt^{-1}$, that using the $\*$-product representation reads
$cpt ( x^\mu \* x^\nu) cpt^{-1}=
cpt_{_{_{}}} x^\mu cpt^{-1}\, \*\, cpt_{_{_{}}} x^\nu cpt^{-1}$. 
{}From here we see that $cpt$ does not act 
on the  $\*$-product. Since $\*\sim
{_{\,}}e^{{i\over 2}{\th}^{\mu\nu}{\stackrel{\leftarrow}
{\partial}_{\!\mu}}{\stackrel{\rightarrow}{\partial}_{\!\nu}}}$
we must have $(i\th^{\mu\nu})^{cpt}\equiv cpt_{_{_{{}}}} (i\th^{\mu\nu}) cpt^{-1}= i\th^{\mu\nu}$
and therefore  $cpt_{_{_{{}}}} (\th^{\mu\nu}) cpt^{-1}= -\th^{\mu\nu}$.
These considerations may be generalized to an $x$ dependent $\th$.
If $\th^{\mu\nu}=C^{\mu\nu\rho}x^\rho$ this means 
$(C^{\mu\nu\rho})^{cpt}= C^{\mu\nu\rho}$;
or we may consider $\th^{\mu\nu}=b^\mu x^\nu- b^\mu x^\mu$
with $(b^{\mu})^{cpt}= b^{\mu}$.

The NCYM actions we have studied are invariant under $CPT$
and therefore under $cpt$ they are invariant iff they are even
in $\th^{\mu\nu}$. Usually this is not the case,
for example $\overline{\widehat{\Psi}}\*{\fmslash\Ah}
\*\widehat{\Psi}$ has a nonvanishing term linear in $\th$.
We conclude that under $cpt$  the NCYM actions we have studied are
not invariant, and $cpt$ is explicitly broken. One could consider 
NCYM actions even in $\th$, for example the most general $SU(2)$ pure 
gauge kinetic term is even in $\th$ because $SU(2)$ has only real 
representations (see Subsection 3.1).
In this case the action is invariant under $cpt$ but $cpt$ is 
{\sl spontaneously} broken because $\th$ itself is not invariant 
under $cpt$. Viceversa if $\th^{\mu\nu}=C^{\mu\nu}_{\rho}x^\rho$ we have
$(C^{\mu\nu}_{\rho})^{cpt}= C^{\mu\nu}_{\rho}$ and with respect to the 
fixed background $C^{\mu\nu}_{\rho}$, $cpt$ is not broken.

\sk
\sk
\noi{\bf Acknowledgements}

\noi This research has been in part supported by a Marie Curie Fellowship
of the European Community programme IHP under contract number MCFI-2000-01982.

\section*{Appendix: Seiberg-Witten maps 
and tensor products}

\paragraph{Gauge parameter:} The most general solution 
(up to redefinition of the ordinary field $A$, and $\Lambda$) to the consistency relation (CR)
\cite{Jurco:2001rq}
\eq
\s[\hat\Lambda_\alpha[A],\hat\Lambda_\beta[A]] + i\delta_\alpha \hat\Lambda_\beta[A]
- i\delta_\beta \hat\Lambda_\alpha[A] =  \hat\Lambda_{[\alpha,\beta]}[A],
\label{consistent}
\en
to first order in $\th$ (which may be non-constant) is
\eq
\hat\Lambda_\Lambda[A] = 
\Lambda + \frac{1}{2} \theta^{ij} \{ A_j , \pp i \Lambda \}_c +
{\mathcal O}(\th^2)~,
\label{generallambda}
\en
where (cf. (\ref{SWLambda}))  
$\hat\Lambda_\alpha[A]=\hat\Lambda[\alpha,A,\th]$,
and for any two matrices $P$ and $Q$
\eq
\{ P , Q \}_c \equiv c P\cdot Q + (1 -c) Q \cdot P
= \frac{1}{2} \{ P, Q \} + (c - \frac{1}{2}) [P, Q] .
\en
The requirement of hermiticity singles out
the preferred choice $c = 1/2$ plus possibly
a purely imaginary function of space-time. The function $c$ also appears
in the following paragraphs.

\paragraph{Covariantizing map:}
The most general differential operator (up to ordinary field redefinitions)
that is a local function of the ordinary
gauge potential $A_i$  and turns a function $f$ into a covariant function
$\mathcal{D}_{[A]}(f)$, with 
\eq
\delta_\Lambda \mathcal{D}_{[A]}(f)
=i\s[\hat\Lambda_\Lambda[A],\mathcal{D}_{[A]}(f)],
\en
to second order in $\theta$ (which may be non-constant) is\footnote{Note:
$\frac{1}{2} \th^{ij} \th^{kl} \{A_j ,A_l\}_c \pp i \pp k
\equiv \frac{1}{2} \th^{ij} \th^{kl} A_j A_l \pp i \pp k $.}
\eq
\mathcal{D}_{[A]} = \mbox{id} + \th^{ij} A_j \pp i 
+ \frac{1}{2} \th^{ij} \{A_j , \pp i (\th^{kl} A_l)\}_c \pp k 
+ \frac{1}{2} \th^{ij} \th^{kl} A_j A_l \pp i \pp k 
+ \frac{1}{2} \th^{ij} \th^{kl} \{F_{il} ,A_j\}_c \pp k + {\mathcal O}(\th^3)~.
\en

\paragraph{NC gauge potential:} Given the covariantizing map
$\mathcal{D}_{[A]}$ the noncommutative gauge
potential $\tilde A^i$ can be read off from the equation
\eq
\mathcal{D}_{[A]}(x^i) = x^i + \tilde A^i.
\en
This gives the following expression for $\tilde A^i$
valid to first order in $\th$ (which may be non-constant)
\eq
\tilde A^i = \th^{il} A_l  
+ \frac{1}{2} \th^{kj} \{A_j , \pp k (\th^{il} A_l)\}_c 
+ \frac{1}{2} \th^{kj} \th^{il} \{F_{kl} ,A_j\}_c + {\mathcal O}(\th^2)~.
\en
For constant non-degenerate $\theta^{il}$ it is more convenient to work 
with $\hat A_l$, where $\tilde A^i = \theta^{il}\hat A_l$.

\paragraph{NC Matter field:}
The most general SW map (up to ordinary field redefinitions) for
(scalar) matter field is
\eq
\widehat \Psi = \psi +\frac{1}{2} \th^{ij} A_j \pp i \psi
+ \frac{1}{4} \th^{ij} (\pp i A_j)\psi - \frac{i}{4} \th^{ij} \{A_i,A_j\}_c \psi
+ {\mathcal O}(\th^2)~.
\en
Note: We have used the classical field redefinition freedom
$\psi  \mapsto \psi + \mu \theta^{ij} F_{ij} \psi$ 
with $\mu = c/4 -1/8$ to obtain
the above formula (where all products of matrices appear within
brackets $\{ , \}_c$).

\subsection*{Tensor products $G \times G'$} 

Consider  the Seiberg-Witten map 
$\mathbf{\hat\Lambda}_{(\Lambda,\Lambda')}[A, A']$
for the gauge parameter 
corresponding to a product $G \times G'$ of gauge groups.
By linearity
in the ordinary gauge parameters $\Lambda$ and $\Lambda'$ we have
\eq
\mathbf{\hat\Lambda}_{(\Lambda,\Lambda')}[A, A'] = 
\mathbf{\hat\Lambda}_{\Lambda}[A, A'] +
\mathbf{\hat\Lambda'}_{\Lambda'}[A, A']. \label{tensorparameter}
\en
The combined gauge parameter $\mathbf{\hat\Lambda}_{(\Lambda,\Lambda')}[A, A']$
should satisfy the CR (\ref{consistent}). This implies that in general
$\mathbf{\hat\Lambda}_{\Lambda}[A, A']$
and $\mathbf{\hat\Lambda'}_{\Lambda'}[A, A']$ satisfy CR's individually and
that there are also new mixed CR's:
\eq
\s[\mathbf{\hat\Lambda}_\Lambda,\mathbf{\hat\Lambda'}_{{\Lambda'}}]
+ i \delta_\Lambda \mathbf{\hat\Lambda'}_{{\Lambda'}} 
- i \delta_{{\Lambda'}}\mathbf{\hat\Lambda}_\Lambda =0
\en
and ditto with $\mathbf{\hat\Lambda} \leftrightarrow \mathbf{\hat\Lambda'}$.
Note that there is no inhomogeneous term on the RHS because
$[\Lambda,{\Lambda'}] = 0 $.
The most general hermitian solution to these equations to order $\th$ is
\begin{eqnarray}
\hat\Lambda_{(\Lambda,_{\!}\Lambda')}[A,A'] &=&
 \Lambda + \Lambda' 
+ \frac{1}{2} \theta^{ij}
\!\Big( \{A_j , \pp i \Lambda\}_c + \{A'_j , \pp i \Lambda'\}_d \!\Big) \nonumber\\
&&+ (1-\frac{\gamma}{2}) \th^{ij} A'_j \pp i \Lambda + \frac{\gamma}{2} 
\th^{ij}\! A_j \pp i \Lambda' + 
{\mathcal O}(\th^2)
\label{productlambda}
\end{eqnarray}
where $\gamma$ is a real function on space-time and $c-1/2$, $d-1/2$ are pure 
imaginary functions.
Comparing (\ref{productlambda}) with (\ref{generallambda}) we see that 
we had to use the freedom of field redefinitions 
$A\rightarrow A+(4-2\gamma)A'$, $A'\rightarrow 2\gamma A'+A$,
to find the gauge parameter for $G \times G'$.
An important special case is given by $\gamma = 1$ and
$d = c$: The corresponding {\bf symmetric solution} can  be obtained
by applying the formula (\ref{generallambda}) to
$\Lambda + \Lambda'$ and $A + A'$. Other interesting special
cases are the {\bf asymetric solutions} for  $\gamma =2$ and $\gamma=0$:
Here one of the two terms in (\ref{tensorparameter}) is given by
the ordinary SW map~(\ref{generallambda}).

{}For the product of (scalar) fields $\Psi\,\Psi'$, where $\Psi$ transforms
under $G$ and $\Psi'$ transforms under $G'$ we can also construct
a SW map $\mathbf{\widehat\Psi}[\Psi,\Psi',A,A']$.\footnote{Note that the
naive choice $\widehat\Psi\star\widehat\Psi'$ does not work,
because the gauge parameter $\widehat\Lambda'$ does not $\star$-commute
with $\widehat\Psi$ in the second term of
$\delta(\widehat\Psi\star\widehat\Psi') =
i\widehat\Lambda\star\widehat\Psi\star\widehat\Psi'
+ i \widehat\Psi\star\widehat\Lambda'\star\widehat\Psi'$.} 
The most general solution to order $\th$ is
\begin{eqnarray}
\lefteqn{\mathbf{\widehat\Psi}[\Psi,\Psi',A,A']
 =   \Psi \Psi' + \th^{\mu\nu} \rho_{\mu\nu} \Psi \Psi'
- \frac{i}{2}(1+\gamma)\th^{\mu\nu} \pp\nu \Psi\pp\mu \Psi'
+ \frac{1}{2} \th^{\mu\nu} (A'_\nu + \gamma A_\nu)\Psi \pp\mu\Psi'}
 \nonumber\\
&& + \frac{1}{2} \th^{\mu\nu} (A_\nu + (2-\gamma) A'_\nu)\pp\mu\Psi \Psi'
+ \frac{1}{2} \th^{\mu\nu}((1-d)\pp\mu A'_\nu + (1-c)\pp\mu A_\nu) \Psi \Psi'
+{\mathcal O}(\th^2) \:,
\qquad
\end{eqnarray}
where $\gamma$, $c$, $d$ are as in equation
(\ref{productlambda}) and $\rho_{\mu\nu}$ is any function (or  differential operator)
that may depend on the gauge potentials $A$, $A'$
and that satisfies
$\delta_\Lambda (\rho_{\mu\nu} \Psi \Psi') = i \Lambda \rho_{\mu\nu}\Psi \Psi'$
and $\delta_{\Lambda'} (\rho_{\mu\nu} \Psi \Psi') 
= i \Lambda' \rho_{\mu\nu}\Psi \Psi'$.
A possibility is $\rho_{\mu\nu} = \rho_{\mu\nu}(F_{\mu\nu}, F'_{\mu\nu}, x)$.
A similar  somewhat lengthy expression exists also
for the noncommutative gauge potential.

An alternative  strategy for the construction of the SW map
for products of fields can be based on the  {\bf hybrid SW map}
(\ref{hybridswmap}):
The SW map for the product of fields $\Psi$ and $\Psi'$ can
be written as (cf. (\ref{35}))
\eq
\mathbf{\widehat\Psi}[\Psi,\Psi',A,A'] = 
\widehat\Psi ^{^H}[\Psi,A + A',A'] \star \widehat\Psi'[\Psi',A'].
\en
The gauge transformation $\delta\Psi = i\Lambda\Psi$,
$\delta\Psi' = i\Lambda'\Psi'$, $\delta A_\mu = \pp\mu \Lambda
+ i[\Lambda,A_\mu]$, $\delta A'_\mu = \pp\mu \Lambda'
+ i[\Lambda',A'_\mu]$  induces the desired transformation
\eq
\delta \mathbf{\widehat\Psi}[\Psi,\Psi',A,A'] = 
i \widehat\Lambda_{(\Lambda+\Lambda')}[A+A'] 
\star \mathbf{\widehat\Psi}[\Psi,\Psi',A,A'],
\en
where we have used that $\Psi$ and $\Lambda'$ commute.
We see that the version of the hybrid SW map 
under consideration corresponds exactly
to the symmetric solution for the gauge parameter.


\end{document}